\documentclass[12pt]{article}
\usepackage[dvips]{graphicx}
\usepackage[american]{babel}
\usepackage{amsmath}
\newtheorem{theorem}{\bfseries Theorem}
\newtheorem{lemma}{\bfseries Lemma}

\begin{document}
\title{Remarks on statistical aspects of safety analysis of complex systems}
\author{L. P\'al\thanks{e-mail: lpal@rmki.kfki.hu} and
M. Makai\thanks{e-mail: makai@sunserv.kfki.hu} \\
\footnotesize{KFKI Atomic Energy Research Institute H-1525
Budapest 114, POB 49 Hungary}} \date{\footnotesize{August 23,
2002}}

\maketitle

\begin{abstract}
{\footnotesize  We analyze safety problems of complex systems
using the methods of mathematical statistics for testing the
output variables of a code simulating the operation of the system
under consideration when the input variables are uncertain. We
have defined a black box model of the code and derived formulas to
calculate the number of runs needed for a given confidence level
to achieve a preassigned measure of safety. In order to show the
capabilities of different statistical methods, firstly we have
investigated one output variable with unknown and known
distribution functions. The general conclusion has been that the
different methods do not bring about large differences in the
number of runs needed to ensure a given level of safety. Analyzing
the case of several statistically dependent output variables we
have arrived at the conclusion that the testing of the variables
separately may lead to false, safety related decisions with
unforseen consequences. We have advised two methods: the sign test
and the tolerance interval methods for testing more then one
mutually dependent output variables.

{\bf List of key words}: safety analysis, black box model, best
estimate, Bayesian method, quantile test, confidence interval,
sign test, tolerance interval.}

\end{abstract}

\newpage

\section{Introduction}

There are two approaches to safety analysis of large complex
systems. Since the analysis has to demonstrate safety of the
operation under the investigated circumstances, we may scrutinize
a not too realistic but rather unfavorable situation saying that
if that situation is safe then any real situation must be on the
safe side. This approach we call {\em conservatism}.

An alternative approach may attempt to investigate the real
situation and show that no limit violation can occur. In this case
the calculated values should be increased by the possible error
when compared with the safety limit \cite{gand88}. That approach
is called {\em best estimate} which is not a very fortunate but
generally accepted name.

In {\em conservative analysis}, the first problem is in the
selection of the case to be studied. It identifies an overt
attempt to bound the actual expected state hence it should
estimate also the consequences of model uncertainties. How do we
know if a given situation is more conservative than is the other?
It is often impossible to foresee the outcome of a non-linear
process. Another problem may be the interplay between
approximations. It may happen that either of two approximations
leads to conservatism but their simultaneous presence does not.
The conservative approach has been prevalent for a long time,
although today rather the best estimate methods are in the focus.

The main difficulty with {\em best estimate calculation} is in the
complexity of the phenomena involved. (A new material phase may
appear, at a given temperature chemical reactions may take place
producing new material properties, and also producing or removing
heat, the process dynamics is nonlinear etc.) In spite of the
problem's complexity, a best estimate method attempts to solve the
equations describing the involved physical processes as accurately
as our knowledge permits. From {\em licensing viewpoint}, several
key parameters should be selected and compared to the acceptance
criteria.

Best estimate methods are accompanied by an {\em uncertainty
analysis} to learn the uncertainty band of the response
\cite{guba03}. The purpose of the uncertainly evaluation is to
provide assurance that the selected parameters at least with
probability $95$\% or more will be in the {\em acceptance region}
or will not exceed their acceptance level.

The present work is dealing with the {\em code uncertainty} only,
which is rather important constituent of the total uncertainty. We
assume the modeled procedure to start from a known initial state.
All the physical quantities in the model we sort as {\em input,
output, and latent data}. By definition, a datum is input if its
domain is known along with a distribution function associating a
probability with any admitted value. In a model, there are several
constants, which are considered either as input or latent data.
Input, when the given constant is looked upon as a variable in a
given range and a probability is allotted to every possible value.
Distinction between input and latent data is a matter of
engineering judgement. The nature of the distribution may depend
on the determination of the constant. Latent, when we refrain from
analyzing the uncertainties of the constant, temporarily we take
it as a fixed number. A datum not falling into the input or latent
category is called output.

The paper is organized as follows. In Section 2 we define a simple
{\em black box model} linking the output variables to input
variables, while in Section 3 we analyze possibilities and
limitations of several well-known statistical tests for {\em one
output variable} with unknown and known cumulative distribution
function. Special attention is paid to the application of a
slightly new variant of {\em the tolerance interval method}. In
Section 4 we deal with the case of several not independent output
variables by using the advantages of {\em order statistics}, and,
finally the conclusions are summarized in Sections 5 and 6.

The present work focuses on deriving {\em criteria for safe
operation} when {\em the output variables are fluctuating as a
result of randomness of input variables}, and intends to give some
help in practical applications. In the sequel we follow the
notation used in the classical handbook of statistics by M.G.
Kendall and A. Stuart \cite{kendall79}.

\section{Black box model}

Let us consider a system as complex as a nuclear power plant, or
an oil refinery plant for instance. Assume we have a model
describing that system, and that model enables us to calculate
physical parameters characterizing the system at arbitrary instant
$t$. Let $n$ be the number of technologically important variables.
In the frame of the model, {\em the operation of the system is
considered safe} if all calculated variables belong to a given set
of intervals
\[{\mathcal V}_{T} = \left\{\left[L_T^{(j)}, U_T^{(j)}\right],
\;\;\; j = 1, \ldots, n \right\} \] determined by the technology.

In order not to be set back by the complexity of the problem, we
suggest a simple {\em black box} model, in which output variables
are linked to input variables. That link can be a {\em computer
code} that transforms vector $\vec{x} \in {\mathcal X}$, the input
variables, into a vector $\vec{y}(t) \in {\mathcal Y}$, the output
variables. Here ${\mathcal X}$ and ${\mathcal Y}$ are sets of all
possible values of $\vec{x}$ and $\vec{y}(t)$, respectively. In
general, the dimension of $\vec{x}$, i.e. the number of input
variables is not the same as the dimension of $\vec{y}(t)$, i.e.
the number of output variables. Every data that enters into the
model is treated as an input variable, hence we do not distinguish
parameters. The model is an explicit relationship between input
$\vec{x}$ and output $\vec{y}$:
\begin{equation} \label{1}
\vec{y}(t) \Leftarrow \hat{\mathcal C}(t)\vec{x},
\end{equation}
where $\hat{\mathcal C}(t)$ is a nonlinear operator that maps
\[ \vec{x} = \left(\begin{array}{ll}
x_{1} \\ x_{2}\\ \vdots \\ x_{h} \end{array} \right)
\;\;\;\;\;\;\mbox{into} \;\;\;\;\;\;
\vec{y}(t) = \left(\begin{array}{ll} y_{1}(t) \\ y_{2}(t)\\ \vdots \\
y_{n}(t) \end{array} \right). \] In practical cases the link
between input and output is very complex hence there is no reason
to anticipate an analytical relationship like $\vec{y}(t) =
\vec{f}(\vec {x},t)$. In the sequel $\hat{\mathcal C}(t)$ is
assumed to be deterministic, in other words once the input has
been fixed, we obtain the same output within the computation
accuracy for each run. At the same time, if the input vector
fluctuates according to distribution laws simulating possible
variations of the technology, or, reflecting uncertainty of some
parameters of the model then the output parameters also fluctuate
in repeated runs.

\begin{figure}[ht!]
\protect \centering{
\includegraphics[height=7cm, width=11cm]{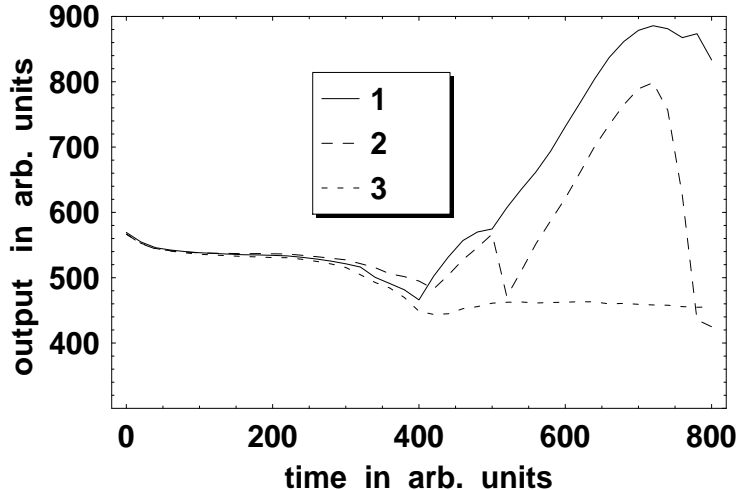}} \protect
\caption{\footnotesize{Influence of the random input on the time
dependence of one output variable in three independent runs.}}
\label{fig1}
\end{figure}

We present an illustration of how random input may influence an
output variable, see Fig. $1$. We used the thermohydraulic code
ATHLET \cite{burw98} to generate several output variables for a
simple experimental setup, but in Fig. $1$ we presented only one
output variable as function of time for three independent runs. It
is obvious that in this case the above given criterion for safe
operation of the system needs to be changed because there is no
guarantee that a new run after a successful run will also be
successful.

We call a state $\vec {x}_{0}$ {\em nominal}, if all the input
parameters take their respective expectation value, i.e.
$\vec{x}_{0} = {\bf E}\{\vec{x}\}$. We can perform a calculation
in the nominal state to get the corresponding output $\vec {y}_{0}
= \hat{\mathcal C}(t) \vec{x}_{0}$. Usually the state
$\vec{x}_{0}$ is called safe if $\vec{y}_{0}$ is in the safety
envelope ${\mathcal V}_{T}$. However, we need a more stringent
definition: {\em state $\vec {x}$ is called safe if $\vec{y}$ is
in the safety envelope ${\mathcal V}_{T}$ for every} $\vec{x} \in
{\mathcal X}$.

Here we should make three remarks. (i) ${\mathcal X}$ may be an
infinite interval when at least one of the input variables is of
normal distribution. In practical calculations such variables are
confined to a finite interval by engineering judgement. (ii) We
check that statement by a given, finite number of calculations
\cite{glser94} with input from ${\mathcal X}$. If there is a value
outside the safety envelop ${\mathcal V}_{T}$ the state $\vec{x}$
is unsafe independently of the fact that the nominal state $\vec
{x}_{0}$ may be safe. (iii) Even if every calculated output is
safe, there is a probability that the state is actually unsafe.

Fixing time $t$ after $N$ runs we obtain $N$ randomly varying
output vectors $\{\vec{y}_{1}(t), \vec{y}_{2}(t), \ldots,
\vec{y}_{N}(t)\}$ which carries information on the fluctuating
input and the code properties. In the next Section we are
considering only one output variable with continuous cumulative
distribution function $G(y) = \int_{-\infty}^{y} g(u)\;du$, and
the time $t$ is taken as fixed and its notation is omitted.

\section{One output variable}

\subsection{Old Bayesian method}

If we carry out $N$ runs with fluctuating input, then we obtain a
sample ${\mathcal S}_{N}=\left\{y_{1}, y_{2}, \ldots, y_{N}
\right\}$ of the random variable $y$ at a fixed time point.
Through technological considerations, we define a fix {\em
acceptance} and a fix {\em rejection interval} to variable $y$.
Let the acceptance interval be ${\mathcal H}_{a} = [L_{T},
U_{T}]$, and ${\mathcal H}_{r} = \overline{\left[L_{T},
U_{T}\right]} = (-\infty, L_{T}) \cup (U_{T}, +\infty)$ the
rejection interval.~\footnote{In many practically important cases
$L_{T} = -\infty$, and so ${\mathcal H}_{a} = (-\infty, U_{T}]$
and ${\mathcal H}_{r} = (U_{T}, +\infty)$.}

The probability
\[ {\mathcal P}\{y \in {\mathcal H}_{a}\} =
\int_{{\mathcal H}_{a}} g(u)\;du = w \] that an observed $y$ lays
in the acceptance interval is not known. Knowing however that $k$
elements of the sample ${\mathcal S}_{N}$ are in the acceptance
interval ${\mathcal H}_{a}$, then utilizing Bayes' theorem,
without knowing $g(u)$, we can claim that
\[ \beta(\omega\vert N,k) = \]
\begin{equation} \label{2}
= \frac{\int_{\omega}^{1} u^{k}\;(1-u)^{N-k}\;du}{\int_{0}^{1}
u^{k}\;(1-u)^{N-k}\;du} = \sum_{j=0}^{k}
\binom{N+1}{j}(1-\omega)^{j}\;\omega^{N + 1 - j} =
\beta(\omega\vert N,k)
\end{equation}
is the probability that the unknown acceptance probability $w$ is
greater than a prescribed $\omega$. The proof of the mentioned
theorem is available in textbooks\footnote{ P\'{a}l. L.:
Fundamentals of probability Theory and Statistics, vol. I.-II.,
109-113, Budapest, Akad\'{e}miai Kiad\'{o}, Budapest (1995), in
Hungarian.} hence we omit it here. We wish to point out the
expression
\begin{equation} \label{3}
\beta(\omega\vert N, 0) = 1 - \omega^{N + 1},
\end{equation}
which shows convincingly that even when the whole sample
${\mathcal S}_{N}$ consists of elements to be accepted, we can
state only that $w \geq \omega$ with probability $1 - \omega^{N +
1}$. If one element in the sample ${\mathcal S}_{N}$ is in the
rejection interval, then we have
\begin{equation} \label{4}
\beta(\omega\vert N, 1) = 1 - \omega^{N + 1} - (N +
1)\;(1-\omega)\omega^{N}.
\end{equation}

Using (\ref{2}), one can easily determine the allowed number of
rejections in a sample of $N$ elements so that the unknown
probability of the acceptance $w$ to be larger than the prescribed
limit $\omega$ with a given probability $\beta(\omega\vert N, k)
\geq \alpha$. It can be agreed on that a system is safe if it is
almost certain ($0 << \alpha \leq 1$) that the unknown probability
of the acceptance $w$ is larger than a prescribed $\omega$.

\newpage

{\bf Table I.} {\footnotesize Number of observations $N$ at which
$w \geq \omega$ with probability $\beta(\omega\vert N,k) \geq
\alpha$ for several values of $\alpha, \omega$, and the number of
rejected values $N-k$.}

\vspace{0.3cm}

\begin{center}
\begin{tabular}{|c|c|c|c|c|} \hline
$\alpha$  & $\omega$  & $N-k=0$ & $N-k=1$ & $N-k=2$ \\ \hline
     & 0.90 &   21  &   31  &   51 \\
0.90 & 0.95 &   44  &   75  &  104 \\
     & 0.99 &  228  &  387  &  530 \\ \hline
     & 0.90 &   27  &   45  &   60 \\
0.95 & 0.95 &   57  &   92  &  123 \\
     & 0.99 &  297  &  472  &  626 \\ \hline
     & 0.90 &   43  &   63  &   80 \\
0.99 & 0.95 &   89  &  129  &  164 \\
     & 0.99 &  457  &  660  &  836 \\ \hline
\end{tabular}
\end{center}

\vspace{0.5cm}

For example, we read out from Table I that if all the 297 observed
values were acceptable, i.e. there was not a single value to be
rejected, then, larger than $95${\%} is the probability that $w
\geq 0.99$, i.e. the proportion of rejected observations in any
sample will be not larger than $0.01$. The more observations we
have, with the higher probability we can state that the
investigated system is safe, and the higher is the lower level
$\omega$ for the unknown acceptance probability $w$.

\subsection{Distribution free confidence interval for quantile}

Assume again the cumulative distribution function $G(y)$ of the
output variable $y$ to be unknown but continuous and strictly
increasing. Denote by $Q_{\gamma}$ the $\gamma$-quantile of
$G(y)$, i.e the value satisfying the equation
\[ G(Q_{\gamma}) = \int_{-\infty}^{Q_{\gamma}} dG(y) = \gamma. \]
Clearly, the interval $(-\infty, Q_{\gamma}]$ covers the
proportion $\gamma$ of the distribution $G(y)$. Since $G(y)$ is
continuous and strictly increasing~\footnote{If $G(y)$ is a
continuous and not decreasing function, then $Q_{\gamma} =
\inf\{y:\;G(y) \geq \gamma\}$.} one can write
\[ Q_{\gamma} = G^{-1}(\gamma).  \]
It is to mention that the {\em point estimate of $Q_{\gamma}$} is
that element of the ordered sample the index $k$ of which is the
nearest integer to $N\gamma$.

\subsubsection{Two-tailed test}

Carrying out $N$ independent runs, we get a sample ${\mathcal
S}_{N} = \{y_{1}, \ldots, y_{N}\}$. Arrange the sample elements in
increasing order,~\footnote{The probability that equal values
occur is zero.} and denote by $y(k)$ the $k$th of ordered
elements; hence we have
\[ y(1) < y(2) < \cdots < y(r) < \cdots < y(s) < \cdots < y(N), \]
and by definition $y(0)=-\infty$, while $y(N+1)=+\infty$. As known
the joint density function of random variables
\[ z(r) = G[y(r)] \;\;\;\;\;\; \mbox{and} \;\;\;\;\;\; z(s) =
G[y(s)], \] where $r$ and $s > r$ are positive integers from $\{1,
2, \ldots, N\}$ is given by
\[ g_{r,s}(u, v) = \frac{u^{r-1}\;(v-u)^{s-r-1}\;(1-v)^{N-s}}{B(r,
s-r)\; B(s, N-s+1)}, \]
\[ 0 \leq u \leq v \leq 1. \] Here $B(j, k)$ is the Euler beta
function.

\begin{theorem}
\label{tr1} If $r$ and $s$ positive integers satisfying the
inequality $0 < r < (N+1)\gamma < s \leq N$, then the random
interval $[y(r), y(s)]$ covers the unknown $\gamma$-quantile
$Q_{\gamma}$ with probability
\[ \beta = {\mathcal P}\{y(r) \leq Q_{\gamma} \leq y(s)\} = \]
\begin{equation} \label{5}
= I(1-\gamma, N-s+1, s) - I(1-\gamma, N-r+1, r)
\end{equation}
where \[ I(c, j, k) = \frac{B(c, j, k)}{B(j,k)}\] is the
regularized incomplete beta function for non-singular cases.
\end{theorem}

The proof of the theorem is simple and it can be find in the
Appendix I. One can see that the confidence level $\beta$ for the
the random interval $[y(r), y(s)]$ does not depend on $G(y)$, in
other words, {\em the confidence interval for the unknown
$Q_{\gamma}$ is distribution free.}

Clearly, there are many different confidence intervals covering
$Q_{\gamma}$ with a prescribed probability $\beta$. We have to
chose the shortest interval by using the following procedure:
\begin{itemize}
\item from the ordered sample determine the integer $q =
[(N+1)\gamma]$ due to the point estimate $\tilde{Q}_{\gamma} =
y(q)$ of the $\gamma$-quantile $Q_{\gamma}$, \item calculate the
confidence level $\beta$ step by step for intervals defined by
integer pairs $[r_{j}, s_{k}]$ where $r_{j} = q - j, \;\; j = 1,
2, \ldots q-1$ and $s_{k} = q + k, \;\; k = 1, 2, \ldots, N-k$,
respectively, until the prescribed value of $\beta$ is reached
provided that it is possible at the sample size $N$ that we have,
\item if the prescribed $\beta$ could not be reached, then the
sample size should have been increased.
\end{itemize}

{\em When the confidence interval $[y(r_{j}), y(s_{k})]$ covering
the $\gamma$-quantile of the unknown distribution $G(y)$ at
prescribed confidence level $\beta$ is a part of the interval
$[L_{T}, U_{T}]$ defined by technology, then the system can be
qualified safe at level $(\beta\vert \gamma)$.}

\vspace{0.3cm}

{\bf Table II.} {\footnotesize Confidence levels $\beta$ for
confidence intervals covering the unknown quantile  $Q_{0.9}$ in
the case of sample size $N=100$. (The point estimate of $Q_{0.9}$
is equal to $\tilde{Q}_{0.9}=y(90)$).}

\vspace{0.2cm}

\begin{center} {\footnotesize
\begin{tabular}{|c|c|c|c|c|c|c|} \hline
$r\backslash s$ & 95  & 96 & 97 & 98 & 99 & 100 \\ \hline
 89 & 0.6455 & 0.6793 & 0.6952 & 0.7011 & 0.7027 & 0.7030 \\ \hline
 88 & 0.7442 & 0.7781 & 0.7940 & 0.7999 & 0.8015 & 0.8018 \\ \hline
 87 & 0.8185 & 0.8524 & 0.8683 & 0.8742 & 0.8758 & 0.8761 \\ \hline
 86 & 0.8699 & 0.9037 & 0.9196 & 0.9255 & 0.9271 & 0.9274 \\ \hline
 85 & 0.9025 & 0.9364 & 0.9523 & 0.9582 & 0.9598 & 0.9601 \\ \hline
 84 & 0.9218 & 0.9557 & 0.9716 & 0.9775 & 0.9791 & 0.9794 \\ \hline
 83 & 0.9324 & 0.9663 & 0.9822 & 0.9880 & 0.9897 & 0.9900 \\ \hline
 82 & 0.9378 & 0.9717 & 0.9876 & 0.9935 & 0.9951 & 0.9954 \\ \hline
 81 & 0.9404 & 0.9743 & 0.9902 & 0.9961 & 0.9977 & 0.9980 \\ \hline
 80 & 0.9416 & 0.9755 & 0.9914 & 0.9972 & 0.9989 & 0.9992 \\ \hline
\end{tabular}}
\end{center}

\vspace{0.3cm}

In Table II we see that, for example, the confidence interval
$[y(85), y(97)]$ defined by elements $y(85)$ and $y(97)$ of the
ordered sample of size $N=100$ covers the quantile $Q_{0.9}$ of
the unknown distribution of the output variable $y$ with
probability (on confidence level) $\beta = 0.9523$. In other
words, having $N=100$ observations for the output variable $y$ we
can state with probability $\beta = 0.9523$ that $y[85] < Q_{0.9}
< y[97]$, i.e. the upper limit of the interval $(-\infty,
Q_{0.9}]$ containing $90$\% of the unknown distribution $G(y)$ is
covered by $[y(85), y(97)]$ on confidence level $\beta = 0.9523$.
If $[y(85), y(97)] \subseteq [L_{T}, U_{T}]$, then {\em the system
is safe, but only on the level $(0.9523\vert 0.9)$.}

When we need stronger criteria of safety, then we have to find
confidence intervals covering quantiles $Q_{0.95}$ or $Q_{0.99}$
with probability near the unity. As seen in Tables III and IV the
sample size $N$ should be greatly increased. For example, if we
would like to construct a confidence interval for the quantile
$Q_{0.99}$ at the level of $\beta = 0.9467$ we need sample with $N
\approx 700$ elements. The production of such a large sample for
even one output variable of complex systems is very expensive, and
at the same time, there is no guarantee that the relation $[y(r),
y(s)] \subseteq [L_{T}, U_{T}]$ will be always satisfied,
especially when the distribution is asymmetric.

\vspace{0.3cm}

{\bf Table III.} {\footnotesize Confidence levels $\beta$ for
confidence intervals covering the $Q_{0.95}$ unknown quantile in
the case of sample size $N=150$. (The point estimate of $Q_{0.95}$
is equal to $\tilde{Q}_{0.95}=y(143)$).}

\vspace{0.2cm}

\begin{center} {\footnotesize
\begin{tabular}{|c|c|c|c|c|c|c|c|} \hline
$r\backslash s$  & 144  & 1456 & 146 & 147 & 148 & 149 & 150 \\
\hline 142 & 0.2909 & 0.4293 & 0.5382 & 0.6090 & 0.6456 & 0.6597 &
0.6633 \\ \hline 141 & 0.4080 & 0.5464 & 0.6553 & 0.7261 & 0.7627 & 0.7768 & 0.7804 \\
\hline 140 & 0.4949 & 0.6333 & 0.7422 & 0.8130 & 0.8496 & 0.8637 &
0.8673 \\ \hline 139 & 0.5531 & 0.6916 & 0.8004 & 0.8712 & 0.9078 & 0.9219 & 0.9255 \\
\hline 138 & 0.5886 & 0.7270 & 0.8359 & 0.9067 & 0.9433 & 0.9574 &
0.9610 \\ \hline 137 & 0.6084 & 0.7469 & 0.8557 & 0.9265 & 0.9632 & 0.9773 & 0.9809 \\
\hline 136 & 0.6186 & 0.7571 & 0.8659 & 0.9368 & 0.9734 & 0.9875 &
0.9911 \\ \hline
\end{tabular}}
\end{center}

\vspace{0.3cm}

{\bf Table IV.} {\footnotesize Confidence levels $\beta$ for
confidence intervals covering the $Q_{0.99}$ unknown quantile in
the case of sample size $N=700$. (The point estimate of $Q_{0.99}$
is equal to $\tilde{Q}_{0.99}=y(694)$).}

\vspace{0.2cm}

\begin{center} {\footnotesize
\begin{tabular}{|c|c|c|c|c|c|c|c|} \hline
$r\backslash s$  & 694  & 695 & 696 & 697 & 698 & 699 & 700 \\
\hline 692 & 0.2808 & 0.4303 & 0.5581 & 0.6490 & 0.7007 & 0.7226 &
0.7289
\\  \hline
 691 & 0.3826 & 0.5321 & 0.6599 & 0.7508 & 0.8024 & 0.8244 & 0.8306 \\  \hline
 690 & 0.4536 & 0.6031 & 0.7309 & 0.8218 & 0.8735 & 0.8954 & 0.9017 \\  \hline
 689 & 0.4986 & 0.6481 & 0.7759 & 0.8668 & 0.9185 & 0.9405 & 0.9467 \\  \hline
 688 & 0.5247 & 0.6742 & 0.8020 & 0.8929 & 0.9446 & 0.9666 & 0.9728 \\  \hline
 687 & 0.5387 & 0.6882 & 0.8160 & 0.9069 & 0.9585 & 0.9805 & 0.9867 \\  \hline
 686 & 0.5456 & 0.6951 & 0.8229 & 0.9138 & 0.9655 & 0.9874 & 0.9936 \\
 \hline
\end{tabular}}
\end{center}

\subsubsection{One-tailed test}

In order to declare that a system is operating safely on a given
level, in many practical cases it seems to be enough to know that
the value of a properly selected output variable $y$ with
probability near $1$ is smaller than the value $U_{T}$ prescribed
by technology. In this case we should determine that element
$y(s)$ of the ordered sample which, with probability $\beta$, is
larger than the quantile $Q_{\gamma}$ of the unknown distribution
$G(y)$ of the output variable $y$. It means that the random
interval $(-\infty, y(s)]$ covers the proportion larger than
$\gamma$ of the unknown distribution $G(y)$ of output variable $y$
with probability
\[ \beta = {\mathcal P}\{y(s) > Q_{\gamma}\}. \]
In order to determine this probability we should substitute $r =
0$ into Eq. (\ref{5}), since according to our definition $y(0) =
-\infty$. We obtain that
\begin{equation} \label{6}
\beta = I(1-\gamma, N-s+1, s) = \sum_{j=0}^{s-1}\binom{N}{j}
\gamma^{j}\;(1-\gamma)^{N-j},
\end{equation}
where $I(c, j, k)$ is the regularized incomplete beta function for
non-singular cases.~\footnote{This equation can be easily derived
directly. It is obvious that
\[ \beta = {\mathcal P}\{y(s) > Q_{\gamma}\} =
{\mathcal P}\{y(s) > G^{-1}(\gamma)\} = {\mathcal P}\{G[y(s)] >
\gamma\}, \] and since the probability density function of the
random variable $z(s) = G[y(s)]$ is nothing else than
\[ g_{s}(u) = \frac{u^{s-1}\;(1-u)^{N-s}}{B(s, N-s+1)}, \]
so we can write immediately that
\[ \beta = \frac{1}{B(s, N-s+1)}\;\int_{\gamma}^{1}
u^{s-1}\;(1-u)^{N-s}\;du = I(1-\gamma, N-s+1, s) = \]
\[ = 1 - I(\gamma, s, N-s+1), \] and this nothing else than (\ref{6}).}
If $y(s)$ is smaller than $U_{T}$, then we can state: {\em the
system is safe at the level $(\beta\vert \gamma)$.}

If $s=N$, i.e. if the largest element of the sample is chosen as
upper limit of the random interval, then one obtains the
well-known formula:
\begin{equation}
\label{7} \beta  = 1 - \gamma^{N}.
\end{equation}
Since in the engineering practice one can find misinterpretations
it is not superfluous to underline the just proven notion of this
formula: $\beta$ is the probability that the largest value $y(N)$
of a sample consisting of $N$ observations  is greater than the
$\gamma$ quantile of the unknown distribution of the output
variable $y$. This statement can be formulated also as follows:
$\beta$ is the probability that the interval $(-\infty, y(N)]$
covers the proportion larger than $\gamma$ of the unknown
distribution $G(y)$ of the output variable $y$.

If $s=N-1$, i.e. if the $(N-1)$-th element of the ordered sample
is chosen as upper limit, then we get from (\ref{6}) the following
formula:
\begin{equation}
\label{8} \beta = 1 - \gamma^{N} - N (1-\gamma)\;\gamma^{N-1},
\end{equation}
the notion of which is obvious.  Clearly, when $\beta$ and
$\gamma$ are fixed, and the second largest element of the sample
is chosen for upper limit, then the sample size $N$ needed to
reach the level $(\beta\vert \gamma)$ is obviously greater than if
the largest element would have been chosen. For example, let the
certainty level $(0.95\vert 0.95)$, then if the largest element is
chosen, the sample size should be $N_{0}=58$,~\footnote{The root
of Eq. $0.95^{N}-0.05 = 0$ is $N \approx 58.404$, and we are using
the rounded value $N=58$. In engineering practice the value $N=59$
is accepted.} while if the second largest one is applied, the
sample size has to be $N_{1}=93$. However, it is at all not
certain that $y^{(93)}(92) \leq y^{(58)}(58)$. (The superscript
denotes the sample size.)

\begin{figure} [ht!]
\label{fig2} \protect \centering{
\includegraphics[height=6cm, width=9cm]{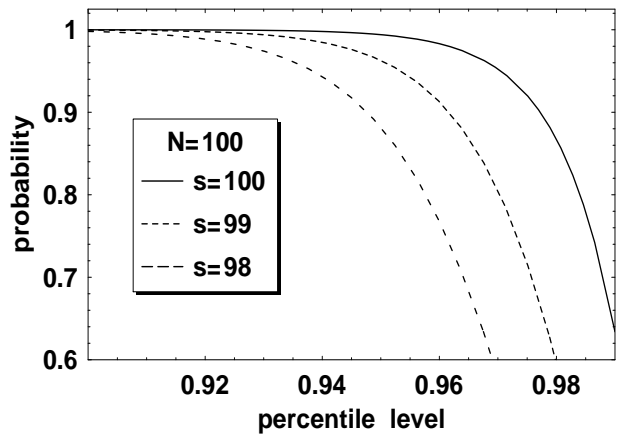}}\protect
\vskip 0.2cm \protect \caption{{\footnotesize Dependence of the
probability $\beta$ on $\gamma = G\left(Q_{p}\right)$ at three
values of $s$.}}
\end{figure}

Figure $2$ shows the dependence of the probability $\beta$ on
$\gamma$ when $N=100$ and $s=100, 99$ and $98$. One can see the
sharp decrease of $\beta$ when the quantile-level $\gamma$
approaches the unity.

\vspace{0.2cm}

{\bf Table V.} {\footnotesize Sample sizes $N_{0}, N_{1}, \ldots,
N_{6}$ for finding elements $y(s),\;\;\; s=N_{0}, N_{1}-1, \ldots,
N_{6}-6$ to be larger than quantiles $Q_{0.90}, Q_{0.95}
\;\mbox{and}\; Q_{0.99}$ of the unknown distribution of the output
variable $y$ with prescribed probabilities $\beta=0.90, 0.95
\;\mbox{and}\;0.99$, respectively.}

\begin{center} {\footnotesize
\begin{tabular}{|c|c|c|c|c|} \hline
$\gamma\backslash \beta$  & 0.90  & 0.95 & 0.99 & $s$ \\
\hline
      &  22 &   28 &   44 &  $N_{0}-0$   \\
      &  37 &   46 &   64 &  $N_{1}-1$   \\
      &  52 &   61 &   81 &  $N_{2}-2$   \\
 0.90 &  65 &   75 &   97 &  $N_{3}-3$   \\
      &  78 &   89 &  112 &  $N_{4}-4$   \\
      &  91 &  102 &  127 &  $N_{5}-5$   \\
      & 103&   115 &  141 &  $N_{6}-6$   \\ \hline
      &  45 &   58 &   90 &  $N_{0}-0$   \\
      &  76 &   93 &  130 &  $N_{1}-1$   \\
      & 105 &  124 &  165 &  $N_{2}-2$   \\
 0.95 & 132 &  153 &  197 &  $N_{3}-3$   \\
      & 158 &  180 &  228 &  $N_{4}-4$   \\
      & 183 &  207 &  258 &  $N_{5}-5$   \\
      & 206 &  234 &  287 &  $N_{6}-6$   \\ \hline
      & 229 &  298 &  458 &  $N_{0}-0$   \\
      & 388 &  473 &  661 &  $N_{1}-1$   \\
      & 531 &  627 &  837 &  $N_{2}-2$   \\
 0.99 & 666 &  773 & 1001 &  $N_{3}-3$   \\
      & 797 &  913 & 1157 &  $N_{4}-4$   \\
      & 925 & 1049 & 1307 &  $N_{5}-5$   \\
      &1051 & 1181 & 1453 &  $N_{6}-6$   \\ \hline
\end{tabular}}
\end{center}

\noindent By fixing the values $\beta$ and $\gamma$ we may
calculate sample sizes $N_{0}, N_{1}, \ldots, N_{k}$ which are
needed for finding elements $y(s), \;\; s=N_{0}, N_{1}-1, \ldots,
N_{k}-k$ such to be larger than the $\gamma$-quantile of the
unknown distribution of the output variable $y$ with prescribed
probability $\beta$. We can see in Table V that for example the
largest element in a sample of size $N=58$ with probability
$\beta=0.95$ is greater than the quantile $Q_{0.95}$ of the
unknown distribution. If $N=234$, then this statement is true for
the element $y(227)$.

\subsubsection{Illustrations}

In order to get a deeper insight into the properties of the just
outlined method, we choose the lognormal distribution with
parameters $m$ and $d$ as the "unknown" distribution $G(y)$. We
note that this distribution arises when many independent random
variables are combined in a multiplicative fashion. The density
function
\[ g(y) = \frac{1}{\sqrt{2 \pi}\;d y}\;\exp\left\{-
\frac{1}{2}\;\left(\frac{\log y -m}{d}\right)^{2}\right\},
\;\;\;\; y \geq 0 \]

\begin{figure}[ht!]
\protect \centering{
\includegraphics[height=6cm, width=10cm]{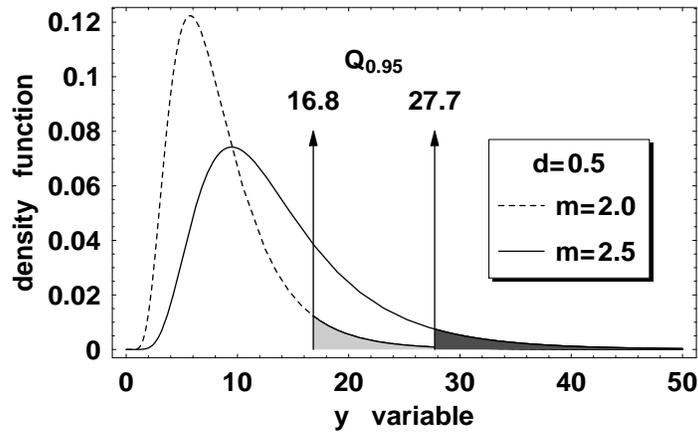}}\protect
\vskip 0.2cm \protect \caption{{\footnotesize Lognormal density
function with parameter values $m=2.0,\;2.5$ and $d = 0.5$. The
vertical arrows indicate the quantile $Q_{0.95}$.}} \label{fig3}
\end{figure}
\noindent can be seen in Fig. $3$ when $m=2.0,\; 2.5$ and $d =
0.5$. The arrows show the quantiles $Q_{0.95} \approx 16.8$
$\;(m=2)\;$ and $Q_{0.95} \approx 27.7$ $\;(m=2.5)$.

By using Monte Carlo simulation let us generate now four sample of
size $N=100$ corresponding to lognormal distribution with
parameters $(m=2.5, \;d=0.5)$, and denote by $A, B, C$ and $D$
these samples. Calculate the point estimates of $0.95$-quantiles
for each of the samples, and determine the shortest two-tailed
confidence intervals which cover with probability $0.95$ the
"unknown" quantile $Q_{0.95}$. In the present case we know that
$Q_{0.95} \approx 16.8$ $\;(m=2)\;$ and $Q_{0.95} \approx 27.7$
$\;(m=2.5)$.

In Fig. $4$ the confidence intervals are shown by vertical
straight lines. Obviously, these intervals are random variables,
hence fluctuate from sample to sample. In the presented example
the sample $D$ is the most unfavorable, because in this case we
can state only that the "unknown" quantile $Q_{0.95}$ is covered
by the interval $[23.29, 53.05]$ with probability larger than
$\beta = 0.95$.

{\bf Table VI.} {\footnotesize Confidence intervals $[y(r), y(s)]$
covering the "unknown" quantile $Q_{0.95}$ with probability
$0.95$.}
\begin{center}
\begin{tabular}{|c|c|c|c|c|} \hline
\mbox{ }  & $A$  & $B$ & $C$ & $D$  \\ \hline $y(r)$ & 22.66 &
25.21 & 22.48 & 23.29 \\  \hline
 $Q_{0.95}$ & {\it 27.73} & {\it 27.73} & {\it  27.73} & {\it 27.73}  \\  \hline
 $y(s)$ & 33.25 & 38.28 & 35.88 & 53.05  \\  \hline
 $(r,s)$ & (91, 100) & (91, 100) & (91, 100) & (91, 100)   \\  \hline
\end{tabular}
\end{center}

\begin{figure}[ht!]
\protect \centering{
\includegraphics[height=8cm, width=12cm]{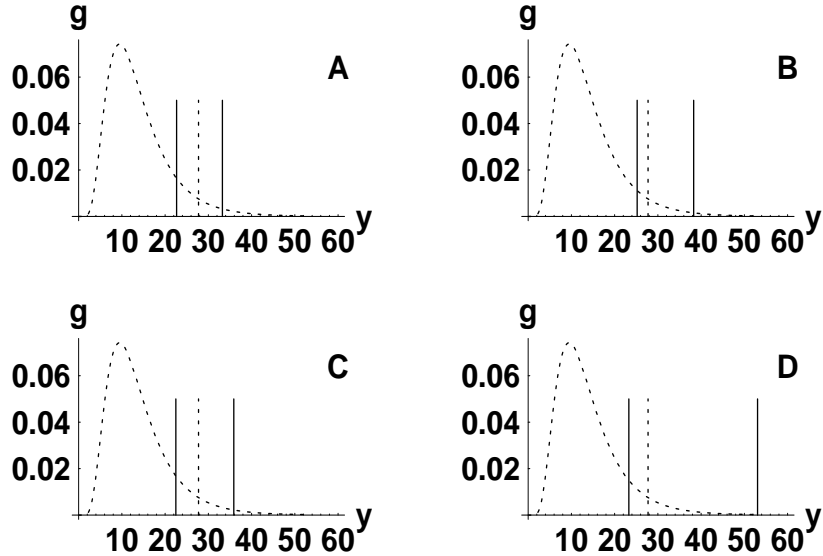}}\protect
\vskip 0.2cm \protect \caption{{\footnotesize Two-sided confidence
intervals denoted by vertical straight lines for samples $A, B, C$
and $D$. The intervals are calculated to be covered the true value
of the quantile $Q_{0.95}$ with probability larger than
$\beta=0.95$. The density function is lognormal with parameters
$m=2.5$ and $d=0.5$. The vertical dashed lines are indicating the
true value of the quantile $Q_{0.95}$.}} \label{fig4}
\end{figure}

\noindent If the upper limit $U_{T}$ determined by technology
would be $U_{T} = 40$, then only three $(A, B, C)$ of four samples
could be regarded safe at the level $(0.95\vert 0.95)$, however,
sample $D$, which is certainly a "rare event", would decrease the
weight of our statement.

As mentioned, in many cases it is enough to know only the element
$y(s), \;\; s = N, N-1, \ldots, N-k$ of the ordered sample of size
$N$ for which the equation
\[ {\mathcal P}\{-\infty \leq Q_{\gamma} < y(s)\} =
{\mathcal P}\{y(s) > Q_{\gamma}\} = \beta \] is valid. The test
based on the interval $(-\infty, y(s)]$ is called {\em one tailed
test}. First, determine the sample size $N$ at which the largest
element of the sample $y(N)$ with probability $\beta$ is greater
than the quantile $Q_{\gamma}$ of the unknown distribution $G(y)$
of the output variable $y$. If $\beta = 0.95$ and $\gamma = 0.95$,
then the largest element has to be chosen out of a sample
containing $N = 58$ elements. Produce a sample of size $N=58$
simulating the lognormal distribution with parameters $m=2.5, \;\;
d=0.5$, and call it {\em basic sample}, denoted by $y_{(b)}$.
Then, repeat randomly the sample production $n$-times, and denote
by $y^{(1)}, y^{(2)}, \ldots, y^{(n)}$ the series of samples. We
are interested in the largest elements $y^{(j)}(58), \;\; j=1,
\ldots, n$ of samples $y^{(j)},\;\; j=1, \ldots, n$.

\begin{figure} [ht!]
\protect \centering{
\includegraphics[height=8cm, width=12cm]{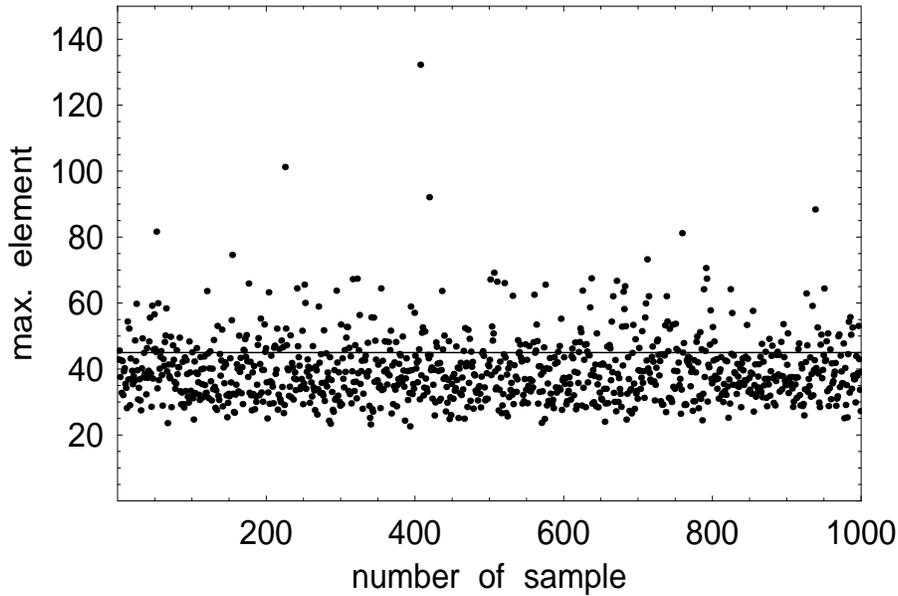}}\protect
\vskip 0.2cm \protect \caption{{\footnotesize Largest elements of
$1000$ samples of size $N=58$. The horizontal line corresponds to
the largest element of the basic sample of size $N=58$. This
element is equal to $y^{(b)}(58) \approx 44.99$.}} \label{fig5}
\end{figure}

Fig. $5$ shows the largest elements of $n=1000$ randomly produced,
independent samples of size $N=58$. The minimal value of the
largest elements is $22.62$, while the maximal value is $132.27$.
One can observe that $224$ largest elements exceed the value
$y^{(b)}(58) \approx 44.99$ which is the largest element of the
{\em basic sample}. However, this surprisingly great number is in
full agreement with the statement that the interval $[0,
y^{(b)}(58)]$ covers the "unknown" $0.95$-quantile with
probability at least $0.95$.

In order to show this, let us introduce the random variable
$\xi_{n}(Q_{\gamma})$ which gives the number of largest elements
being greater than the quantile $Q_{\gamma}$ in $y^{(j)},\;\; j=1,
\ldots, n$ independent samples of size $N$. Since the probability
that the largest element in a given sample is greater than
$Q_{\gamma}$ is nothing else than $1 - \gamma^{N}$, hence, we
conclude that
\[ {\mathcal P}\{\xi_{n}(Q_{\gamma}) = k\} = \binom{n}{k}
\;(1-\gamma^{N})^{k}\;\gamma^{N(n-k)}. \] From this we obtain
immediately that
\[ {\bf E}\{\xi_{n}(Q_{\gamma})\} = n(1-\gamma^{N}) \;\;\;\;\;\;
\mbox{and} \;\;\;\;\;\;  {\bf D}\{\xi_{n}(Q_{\gamma})\} =
\sqrt{n\;\gamma^{N}\;(1-\gamma^{N})}.\] As known, if $n$ and $k$
are sufficiently large, then the distribution of the random
variable
\[ \chi_{n}(Q_{\gamma}) = \frac{\xi_{n}(Q_{\gamma}) -
{\bf E}\{\xi_{n}(Q_{\gamma})\}}{{\bf D}\{\xi_{n}(Q_{\gamma})\}} \]
is  approximately standard normal, hence we can write that
\[ w = {\mathcal P}\left\{\vert \chi_{n}(Q_{\gamma}) \vert \leq
\lambda\right\} = \] \[ = {\mathcal P}\left\{{\bf
E}\{\xi_{n}(Q_{\gamma})\} - \lambda\;{\bf
D}\{\xi_{n}(Q_{\gamma})\} \leq \xi_{n}(Q_{\gamma}) \leq {\bf
E}\{\xi_{n}(Q_{\gamma})\} + \lambda\;{\bf
D}\{\xi_{n}(Q_{\gamma})\}\right\}, \] where $\lambda$ is the root
of Eq.
\[ \frac{1}{\sqrt{2\pi}}\;\int_{-\infty}^{\lambda}
e^{-u^{2}/2}\;du = \frac{1+w}{2}. \] It means that the inequality
\[ {\bf E}\{\xi_{n}(Q_{\gamma})\} - \lambda\;{\bf
D}\{\xi_{n}(Q_{\gamma})\} \leq \xi_{n}(Q_{\gamma}) \leq {\bf
E}\{\xi_{n}(Q_{\gamma})\} + \lambda\;{\bf
D}\{\xi_{n}(Q_{\gamma})\}\] is valid with probability $w$.

If $n=1000, \;\; N=58, \gamma = 0.95$ and $w =0.95$, then we
obtain the values ${\bf E}\{\xi_{n}(Q_{\gamma})\} = 950,\;\; {\bf
D}\{\xi_{n}(Q_{\gamma})\} \approx 6.96$ and $\lambda \approx
1.96$,  hence we can state with probability $0.95$ that
\[ 936 < \xi_{1000}(Q_{0.95}) < 964. \]
If we count the number of largest elements $y^{(j)}(58), \;\;
j=1,\ldots,1000$ exceeding $Q_{0.95}$ that we know in this example
($Q_{0.95} \approx 27,728$), we obtain the value $949$ that is
indeed inside of the interval $[936, 964]$.

In spite of this "nice" agreement we have to underline that the
requirement of safety, for instance, at the level $(0.95\vert
0.95)$ does not exclude the appearance of "rare events" such as
exceeding the technological limit $U_{T}$. Therefore, we advice
{\em stronger requirements of safety} the fulfillment of which, of
course, is much more expensive.

\subsection{Method based on sign test}

Assume again the cumulative distribution function $G(y)$ of the
output variable $y$ to be continuous but unknown. Let ${\mathcal
S}_{N} = \{y_{1}, \ldots, y_{N}\}$ be a sample containing the
values of $N$ observations. Define the function
\[ \Delta(x) = \left\{ \begin{array}{ll}
1, & \mbox{if $x > 0$,} \\
\mbox{} & \mbox{} \\
0, & \mbox{if $x < 0$,} \end{array} \right. \] and introduce the
statistical function
\begin{equation}
\label{9} z_{N} = \sum_{j=1}^{N} \Delta(U_{T} - y_{j}).
\end{equation}
which gives the number of sample elements smaller than $U_{T}$.
Criteria based on this statistical function are used to be named
{\em sign criteria} because $z_{N}$ counts only the positive
differences $U_{T} - y_{j}, \;\; j = 1, \ldots, N$. Since we
assumed that $G(y)$ is continuous, hence the probability of the
event $\{U_{T} - y = 0\}$ is zero.

Obvious that $z_{N}$ has {\em binomial distribution} since $z_{N}$
is nothing else than the sum of $N$ independent random variables
with values either $0$ or $1$. By using the notation
\begin{equation}
\label{10} {\mathcal P}\{\Delta(U_{T}-y) = 1\} = {\mathcal P}\{y
\leq U_{T}\} = p,
\end{equation}
we can write
\begin{equation} \label{11}
{\mathcal P}\{ z_{N} = j\} = \binom{N}{j} p^{j}\;(1-p)^{N-j},
\end{equation}
\[ j = 0, 1, \ldots, N. \]

The task is very simple. Assume that we have a sample of size $N$
and for this sample $z_{N} = k \leq N$. We should determine a
confidence interval $[\gamma_{L}(k), \gamma_{U}(k)]$ which covers
the value $p$ with a prescribed probability $\beta$.  The unknown
$p$ defined by (\ref{10}) is nothing else than the probability
that the output variable $y$ is not larger than the technological
limit $U_{T}$. When the lower confidence limit $\gamma_{L}(k)$ is
near the unity, then, since $\gamma_{L}(k) < p$, we can state at
least with probability $\beta$ that the chance of finding the
output variable $y$ smaller than $U_{T}$ is also near the unity,
and so the system operation can be regarded safe at the level
$[\beta\vert \gamma_{L}(k)]$.

\subsubsection{Approximate calculation}

If the sample size $N > 50$, then the random variable
\[ \frac{k - Np}{\sqrt{Np\;(1 - p)}} = \zeta_{k} \]
has approximately {\em standard normal distribution}, where $k$ is
the number of sample elements not larger than $U_{T}$. Let $\beta$
be the confidence level, then we can write that
\[ {\mathcal P}\{\vert \zeta_{k} \vert \leq u_{\beta}\} = {\mathcal
P}\left\{\frac{\vert k - Np\vert}{\sqrt{Np\;(1 - p)}} \leq
u_{\beta}\right\} = 2\Phi(u_{\beta}) - 1 = \beta, \] where
$\Phi(x)$ is the standard normal distribution function. This
equation can be rewritten~\footnote{The following elementary
considerations can be found in any textbook for statistics, e.g.
\cite{lpal95}.} in the following form:
\[ {\mathcal P}\{\vert \zeta_{k} \vert \leq u_{\beta}\} =
{\mathcal P}\{\zeta_{k}^{2}\leq u_{\beta}^{2}\} = \]
\begin{equation} \label{12}
= {\mathcal P}\{ (N+u_{\beta}^{2})(p-\gamma_{L})(p-\gamma_{U})
\leq 0\} = \beta,
\end{equation}
where
\begin{equation}\label{13}
\gamma_{L} = \gamma_{L}(k, u_{\beta}) =
\frac{1}{N+u_{\beta}^{2}}\;\left[k + \frac{1}{2}u_{\beta}^{2} -
u_{\beta}\sqrt{k(1-k/N) + u_{\beta}^{2}/4}\right],
\end{equation}
and
\begin{equation}\label{14}
\gamma_{U} = \gamma_{U}(k, u_{\beta}) =
\frac{1}{N+u_{\beta}^{2}}\;\left[k + \frac{1}{2}u_{\beta}^{2} +
u_{\beta}\sqrt{k(1-k/N) + u_{\beta}^{2}/4}\right].
\end{equation}
It is obvious that $[p-\gamma_{L}(k, u_{\beta})][p-\gamma_{U}(k,
u_{\beta})] \leq 0$ is fulfilled only, if
\[ \gamma_{L}(k, u_{\beta}) \leq p \leq \gamma_{U}(k, u_{\beta}), \]
and therefore
\begin{equation}\label{15}
{\mathcal P}\{\vert \zeta_{k} \vert \leq u_{\beta}\} = {\mathcal
P}\{\gamma_{L}(k, u_{\beta})\leq p \leq \gamma_{U}(k, u_{\beta})\}
= \beta
\end{equation}
where $u_{\beta}$ is the root of Eq.
\[ \Phi(u_{\beta}) = \frac{1}{2}(1 + \beta). \]
This equation shows clearly that the interval $[\gamma_{L}(k,
u_{\beta}),\; \gamma_{U}(k, u_{\beta})]$ covers the unknown $p$
with probability $\beta$.

In many cases we do not need the restriction due to the upper
confidence limit. We want to know only the probability of the
event $\{\gamma_{L}(k, v_{\beta}) \leq p\}$. Since $\zeta_{k}$ at
fixed $k$ is a decreasing function of $p$, the events $\{\zeta_{k}
\leq v_{\beta}\}$ and  $\{\gamma_{L}(k, v_{\beta}) \leq p\}$ are
equivalent, and so we can write
\begin{equation} \label{16}
{\mathcal P}\{ \zeta_{k} \leq v_{\beta}\} = {\mathcal P}\{
\gamma_{L}(k, v_{\beta}) \leq p\} = \Phi(v_{\beta}) = \beta.
\end{equation}
Consequently, the operation of a system can be regarded safe if
the parameter $p$ for all output variables is covered by
$[\gamma_{L}(k, v_{\beta}),\;1]$ with a prescribed probability
$\beta$, provided that $\gamma_{L}(k, v_{\beta})$ is near the
unity.~\footnote{It is obvious that $\gamma_{L}(k, v_{\beta}) \geq
\gamma_{L}(k, u_{\beta})$.} For the sake of simpler notation in
the sequel $\gamma_{L}(k, v_{\beta})$ and $\gamma_{U}(k,
u_{\beta})$ will be denoted by $\gamma_{L}$ and $\gamma_{U}$,
respectively.

The event $\{y \leq U_{T}\}$ belonging to the {\em acceptance
region} of the sample space will be called {\em success}. Now, let
us calculate the number of successes $k$ needed in a sample of
size $N$  to ensure a fixed confidence level $\beta$ and a given
lower confidence limit $\gamma_{L}$.

\vspace{0.3cm}

{\bf Table VII.} \hspace{0.1cm} {\footnotesize Numbers of sample
elements $k$ in samples of size $N=100(10)200$ needed for the
acceptance on level $\beta = \gamma_{L} = 0.95$.  (The approximate
formula $(13)$ has been used for calculations.)}

\begin{center}
\begin{tabular}{|c|c|c|c|c|c|c|c|c|c|c|c|} \hline
$k$ & 99 & 108 & 118 & 128 & 137 & 147 & 157 & 166 & 176 & 185 &
195 \\ \hline $N$ & 100 & 110 & 120 & 130 & 140 & 150 & 160 & 170
& 180 & 190 & 200  \\ \hline
\end{tabular}
\end{center}

\vspace{0.3cm}

In Table VII we see the numbers of successes needed in samples of
size $N=100(10)200$ in order to reach the level $\beta =
\gamma_{L} = 0.95$. The requirement is quite sever: if the sample
size $N=100$ one should have $k=99$ successes!

For illustration of the method the approximate $\gamma_{L} < p$
values have been calculated at confidence levels $\beta =
0.90(0.01)0.99$ when the sample size $N=100$ and the number of
successes $k = 90(1)100$. The results are shown in Table VIII. It
can be seen, for example, that if the event $\{y \geq U_{T}\}$
occurs only once, then it can be stated with probability $\beta =
0.95$ that $\gamma_{L} = 0.9564 < p$. It means that the appearance
of "dangerous" events $\{y \geq U_{T}\}$ is not excluded even if
the level of acceptance is better than $(0.95\vert 0.9564)$.

\vspace{0.3cm}

{\bf Table VIII.} {\footnotesize Approximate $\gamma_{L} < p$
values calculated at confidence levels $\beta = 0.90(0.01)0.99$
for numbers of success $k = 90(1)100$. Sample size $N=100$.}

\begin{center}
{\footnotesize
\begin{tabular}{|c|c|c|c|c|c|c|} \hline
$k$ $\backslash$ $\beta$ & 0.90 & 0.91 & 0.92 & 0.93 & 0.94 & 0.95
\\ \hline 90 & 0.8549 & 0.85245 & 0.8498 & 0.8469 & 0.8435 &
0.8396 \\ \hline  91 & 0.8664 & 0.8640 & 0.8615 & 0.8586 & 0.8553
& 0.8515 \\ \hline 92 & 0.8781 & 0.8758 & 0.8733 & 0.8704 & 0.8672
& 0.8635 \\ \hline 93 & 0.8899 & 0.8877 & 0.8852 & 0.8825 & 0.8794
& 0.8757 \\ \hline 94 & 0.9019 & 0.8997 & 0.8974 & 0.8947 & 0.8917
& 0.8882 \\ \hline 95 & 0.9141 & 0.9120 & 0.9097 & 0.9072 & 0.9043
& 0.9008 \\ \hline 96 & 0.9266 & 0.9246 & 0.9224 & 0.9200 & 0.9171
& 0.9138 \\ \hline 97 & 0.9394 & 0.9376 & 0.9355 & 0.9331 & 0.9304
& 0.9273 \\ \hline 98 & 0.9528 & 0.9511 & 0.9491 & 0.9469 & 0.9444
& 0.9414 \\ \hline 99 & 0.9672 & 0.9655 & 0.9637 & 0.9617 & 0.9593
& {\bf 0.9564} \\ \hline 100   & 0.9838 & 0.9823 & 0.9806 & 0.9787
& 0.9764 & 0.9737 \\ \hline
\end{tabular}}
\end{center}
\begin{center}
{\footnotesize
\begin{tabular}{|c|c|c|c|c|} \hline
$k$ $\backslash$ $\beta$ & 0.96 & 0.97 & 0.982 & 0.99 \\ \hline 90
& 0.8350 & 0.8292 & 0.8213 & 0.8085 \\ \hline 91 & 0.8470 & 0.8413
& 0.8335 & 0.8208 \\ \hline 92 & 0.8591 & 0.8535 & 0.8458 & 0.8333
\\ \hline 93 & 0.8714 & 0.8659 & 0.8584 & 0.8460 \\ \hline 94 &
0.8839 & 0.8786 & 0.8712 & 0.8591 \\ \hline 95 & 0.8967 & 0.8915 &
0.8843 & 0.8724 \\ \hline 96 & 0.9099 & 0.9048 & 0.8978 & 0.8861
\\ \hline 97 & 0.9235 & 0.9186 & 0.9117 & 0.9003 \\ \hline 98 &
0.9377 & 0.9330 & 0.9264 & 0.9152 \\ \hline 99 & 0.9529 & 0.9484 &
0.9420 & 0.9311 \\ \hline 100 & 0.9703 & 0.9658 & 0.9505 & 0.9487
\\ \hline
\end{tabular}}
\end{center}

\subsubsection{Exact calculation}

When the sample size $N$ is smaller than $50$ we cannot apply the
asymptotically valid normal distribution. For the exact
calculation of confidence limits we used a slightly new version of
the method proposed by Clopper and Pearson \cite{pearson34}.

The probability of finding at least $k$ successes from $N$
observations is nothing else than
\begin{equation} \label{17}
S_{k}^{(N)}(p) = \sum_{j=0}^{k} \binom{N}{j} p^{j}\;(1-p)^{N-j},
\end{equation}
where
\[ p = {\mathcal P}\{y \leq U_{T}\}. \]
As known, this formula can be written in the form:
\[ S_{k}^{(N)}(p) = \frac{N!}{k!\;(N-k-1)!}\;\int_{p}^{1}
u^{k}\;(1-u)^{N-k-1}\;du = \]
\begin{equation} \label{18}
= \frac{N!}{k!\;(N-k-1)!}\;\int_{0}^{1-p}
(1-v)^{k}\;v^{N-k-1}\;dv,
\end{equation}
and it is obvious, that $S_{k}^{(N)}(p)$ is a continuous monotone
decreasing function of $p$, since
\[ \frac{dS_{k}^{(N)}(p)}{dp} = - \frac{N!}{k!\;(N-k-1)!}\;p^{k}\;
(1-p)^{N-k-1} < 0. \] Taking into account that
\[ S_{k}^{(N)}(p) = \left\{ \begin{array}{ll}
1, & \mbox{if $p=0$,} \\
\mbox{} & \mbox{} \\
0, & \mbox{if $p=1$,} \end{array} \right. \] it is evident that
$S_{k}^{(N)}(p)$ assumes any values in the interval $[0, 1]$ only
once. Consequently, a $p = p_{\delta}$ value can be determined so
that
\[ S_{k}^{(N)}(p_{\delta}) = \delta, \;\;\;\;\;\; \forall \; 0
< \delta < 1. \] Since $S_{k}^{(N)}(p)$ is a monotone decreasing
function, if $p > p_{\delta}$, then
\[ S_{k}^{(N)}(p) < S_{k}^{(N)}(p_{\delta}) = \delta.\]

\begin{figure} [b!]
\protect \centering{
\includegraphics[height=14cm, width=10cm]{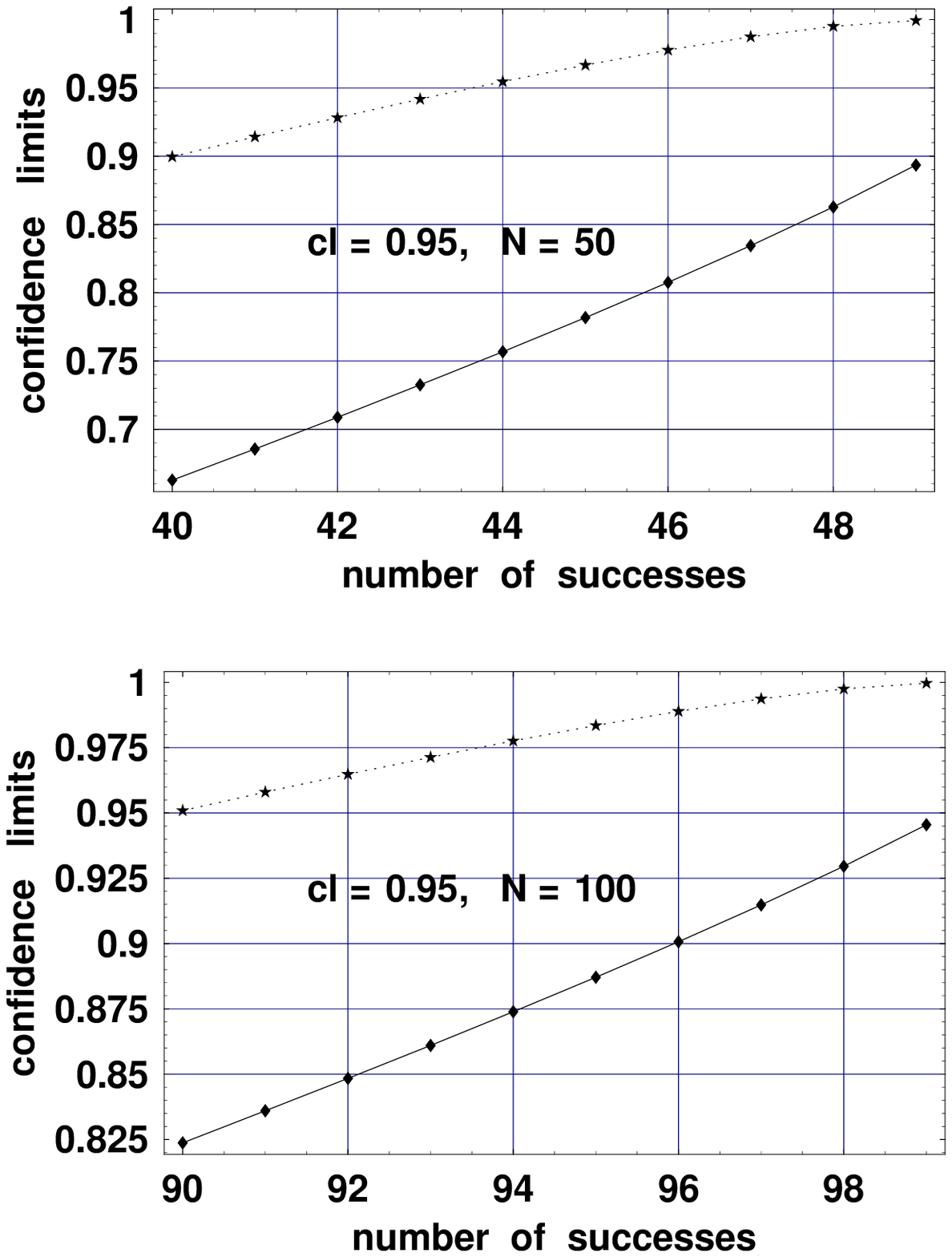}}\protect
\vskip 0.2cm \protect \caption{{\footnotesize Dependence of the
upper and the lower confidence limits on the number of successes
$k$ at confidence level $\beta=cl=0.95$ in cases of sample size
$N=50$ and $100$, respectively.}} \label{fig6}
\end{figure}

Clearly, the function
\begin{equation} \label{19}
R_{k}^{(N)}(p) = 1 - S_{k-1}^{(N)}(p) = \sum_{j=k}^{N}
\binom{N}{j} p^{j}\;(1-p)^{N-j},
\end{equation}
will satisfy the inequality
\[  R_{k}^{(N)}(p) < R_{k}^{(N)}(p_{\delta}) = \delta, \;\;\;\;\;\; \mbox{if}
\;\;\;\;\;\; p < p_{\delta}. \] Fixing the confidence level
$\beta$ one can obtain the upper confidence limit $\gamma_{U}$ for
the unknown parameter $p$ from $S_{k}^{(N)}(\gamma_{U}) \leq
\frac{1}{2}(1-\beta)$, while the lower confidence limit
$\gamma_{L}$ is determined by $R_{k}^{(N)}(\gamma_{L}) \leq
\frac{1}{2}(1-\beta)$. Now one can formulate the statement that
the random interval $[\gamma_{L}, \; \gamma_{U}]$ covers the
un\-known parameter $p$ with probability $\beta$.

\begin{figure} [ht!]
\protect \centering{
\includegraphics[height=6cm, width=9cm]{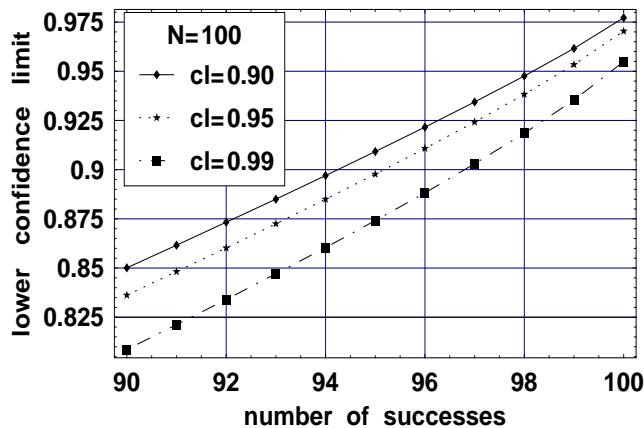}}\protect
\vskip 0.2cm \protect \caption{{\footnotesize Dependence of the
the lower confidence limit on the number  of successes $k$ on
three confidence levels $ \beta = cl = 0.90,\;0.95, \; 0.99$ when
the sample size $N = 100$.}} \label{fig7}
\end{figure}

For the sake of illustration  Fig. \ref{6} shows the dependence of
the upper and the lower confidence limits on the number of
successes $k$ on confidence level $\beta = 0.95$ in cases of
sample size $N=50$ and $100$, respectively. For example, if
$k=98$, i.e. two observations out of $N=100$ are failed, then we
can state with probability $0.95$ that the unknown $p$ is covered
by the interval $[0.9296, \; 0.9975]$.

As mentioned already in many practical situations it suffices  to
know that the interval $[\gamma_{L}, \;1]$ calculated from the
sample of $N$ observations covers the chance of success $p =
{\mathcal P}\{y \leq U_{T}\}$ with prescribed probability $\beta$.
Fig. $7$ shows the dependence of the the lower confidence limit
$\gamma_{L}$ on the number of successes $k$ at three confidence
levels $\beta = cl = 0.90,\;0.95, \; 0.99$ when the sample size
$N=100$.

\vspace{0.3cm}

{\bf Table IX.} {\footnotesize  Lower confidence limits at three
levels when the number of successes $k = 90(1)100$. Sample size
$N=100$.}

\begin{center}
{\footnotesize
\begin{tabular}{|c|c|c|c|c|c|c|} \hline
$\beta$ $\backslash$ $k$ & 90 & 91 & 92 & 93 & 94 & 95 \\ \hline
0.90 & 0.8501 & 0.8616 & 0.8733 & 0.8850 & 0.8970 & 0.9092 \\
\hline 0.95 & 0.8362 & 0.8482 & 0.9602 & 0.9725 & 0.8850 & 0.8977
\\ \hline 0.99 & 0.8086 & 0.8212 & 0.8340 & 0.8471 & 0.8604 &
0.8741 \\ \hline
\end{tabular}}
\end{center}
\begin{center}
{\footnotesize
\begin{tabular}{|c|c|c|c|c|c|} \hline
$\beta$ $\backslash$ $k$ &  96 & 97 & 98 & 99 & 100 \\ \hline 0.90
& 0.9216 & 0.9344 & 0.9476 & 0.9616 & 0.9772 \\ \hline 0.95 &
0.9108 & 0.9242 & 0.9383 & 0.9534 & {\bf 0.9704} \\ \hline 0.99 &
0.8882 & 0.9030 & 0.9185 & 0.9354 & 0.9549 \\ \hline
\end{tabular}}
\end{center}

\vspace{0.3cm}

Table IX contains the $\gamma_{L}$ values plotted in Fig. $7$ for
the mostly used confidence levels provided that the sample size
$N=100$.  It is remarkable that even in that case when $k=100$,
i.e. when all elements of a sample can be found in the acceptance
interval we can state with probability $\beta = 0.95$ only that
the unknown $p$ value is covered by the interval $[0.9704, \; 1]$,
or simply, but not precisely: the $p$ is larger than $0.97$ with
probability $0.95$ One can imagine a number of cases where this
statement is not enough to declare: the operation of the analyzed
system can be regarded safe.

\subsection{Tolerance interval method}

Assume again that we have $N$ independent values $y_{1}, \ldots,
y_{N}$ of the output variable $y$. Let $\gamma$ and $\beta$ be
positive numbers not larger than $1$. Now, we wish to answer the
following question: {\em On the basis of a sample ${\mathcal
S}_{N} = \{y_{1}, \ldots, y_{N}\}$ can we state that a fraction
larger than $\gamma$ of the distribution $G(y)$ lays with
probability $\beta$ in an interval $[L, U] \subseteq [L_{T},
U_{T}]$?}

In order to answer this question, let us construct from the sample
${\mathcal S}_{N}$ two random functions $L = L(y_{1}, \ldots,
y_{N})$ and $U = U(y_{1}, \ldots, y_{N})$, called {\em tolerance
limits}, such that
\begin{equation} \label{20}
{\mathcal P}\{\int_{L}^{U} dG(y) > \gamma\} = \beta.
\end{equation}
We remark that
\begin{equation} \label{21}
\int_{L}^{U} dG(y) = {\mathcal A}(y_{1}, \ldots, y_{N})
\end{equation}
is a random variable, sometimes called {\em probability content},
which measures the proportion of the distribution included in the
random interval $[L,U]$. Probability $\beta $ bears the name {\em
confidence level}. For safe operation it is advisable to specify
the probability content $\gamma$ and the confidence level $\beta$
as large as possible in the interval $(0,1)$.

Having fixed $\beta $ and $\gamma $, from definitions of $L(y_{1},
\ldots, y_{N})$ and $U(y_{1}, \ldots, y_{N})$ it becomes possible
to determine the number of runs $N$. Carrying out $N$ runs, we get
a sample $\{y_{1}, \ldots, y_{N}\}$, from which we can calculate
an appropriate tolerance interval $[L,U]$. If that interval lies
in $[L_{T},U_{T}]$ we declare the operation safe.~\footnote{ Many
authors have discussed the problem of setting tolerance limits for
a distribution on the basis of an observed sample. The pioneering
work was done by S. S. Wilks \cite{wilks42} and by A. Wald
\cite{wald46}.} This program can be easily realized when the
distribution $G(y)$ is known and normal, however, in subsection
$3.4.1$ the problem of distribution free tolerance interval will
be discussed.

\subsubsection{Distribution free tolerance limits}

To solve the problem of setting tolerance limits when nothing is
known about the cumulative distribution function $G(y)$ except
that it is continuous, seems to be not an easy task. Exploiting
advantages of the order statistics, Wilks \cite{wilks42} was the
first who found a satisfactory solution to the problem and
somewhat later Robbins \cite{robb44} published a nice proof that
{\em distribution free tolerance limits can be given only by means
of order statistics}.

It is evident that in the order statistics we are unable to
exploit the total amount of information which is present in the
sample when the distribution function $G(y)$ is unknown.
Consequently, with $\gamma$ and $\beta$ given, we anticipate
either a wider tolerance interval around the sample mean or a
larger sample size to achieve the same tolerance interval as in
the case of known $G(y)$. Not going into details, we give here a
well-known theorem, which is useful in uncertainty and sensitivity
analysis of codes.

\begin{theorem} \label{tr2}
Let $y_{1}, \ldots, y_{N}$ be $N$ independent observations of the
random output $y$. Suppose that nothing is known about the
distribution function $G(y)$ except that it is
continuous.~\footnote{It can be shown that the one-sided
continuity only is needed.} Arrange the values of $y_{1}, \ldots,
y_{N}$ in increasing order,~\footnote{The probability that equal
values occur is zero.} and denote by $y(k)$ the $k$-th  of these
ordered values; hence in particular
\[ y(1) = \min_{1 \leq k \leq N} y_{k}, \;\;\;\;\;\; y(N) = \max_{1
\leq k \leq N} y_{k},\] and by definition $y(0)=-\infty$, while
$y(N+1)=+\infty$. In this case for some positive $\gamma < 1$ and
$\beta < 1$ there can be constructed two random function $L(y_{1},
\ldots, y_{N})$ and $U(y_{1}, \ldots, y_{N})$, called tolerance
limit, such that the probability that
\[ \int_{L}^{U} dG(y) > \gamma \] holds is equal to
\begin{equation} \label{22}
\beta = 1 - I(\gamma, s-r, N-s+r+1) = \sum_{j=0}^{s-r-1}
\binom{N}{j} \gamma^{j}\;(1 - \gamma)^{N-j},
\end{equation}
where
\begin{equation} \label{23}
I(\gamma, j, k) =
\int_{0}^{\gamma}\frac{u^{j-1}\;(1-u)^{k-1}}{B(j, k)}\;du,
\;\;\;\;\;\; B(j, k) = \frac{(j-1)!\;(k-1)!}{(j+k-1)!},
\end{equation}
\[ 0 \leq r < s \leq N, \;\;\;\;\;\; \mbox{and} \;\;\;\;\; L =
y(r), \;\;\;\;\; U = y(s). \]
\end{theorem}

The proof of Theorem 2, which is a simplified version of Wald's
proof, is given in Appendix II.

The selection of tolerance limits $L=y(1)$ and $U=y(N)$ appears to
be expedient in many cases. Substituting $r=1$ and $s=N$ in Eq.
(\ref{22}), we get for the {\em two-sided tolerance interval} the
expression
\begin{equation} \label{24}
\beta = 1 - \gamma^{N} - N(1 - \gamma)\;\gamma^{N-1}.
\end{equation}
Often we are interested solely in the upper tolerance limit
$U=y(N)$ and we call the interval $[y(0),y(N)]$ {\em one-sided
tolerance} interval. Now $r=0$ and $s=N,$ therefore
\begin{equation} \label{25}
\beta = 1 - \gamma^{N}.
\end{equation}
When the lower limit is of interest, we select $[y(1),y(N+1)]$ and
this is also a one-sided tolerance interval. Substituting r=1 and
s=N+1 into expression (\ref{22}), we obtain (\ref{25}) again.

Finally, we make two remarks. Two outputs are considered the same
if their difference is smaller than the round-off error. Therefore
the probability that two runs yield the same output is very small
but not zero. The second remark is that expressions
(\ref{24})-(\ref{25}) may appear as a relationship between two
probabilities $\beta $ and $\gamma $. However, $\gamma$ is not a
probability, which can be seen from the nonsensical interpretation
for $\gamma$ from any of the mentioned expressions. In Table X. we
compiled the probability content $\gamma $ of the tolerance
interval $[y(1), y(N)]$ for $\beta=0.9, 0.95, 0.99$ and
$N=10(10)100(25)300$.

If we are interested in a tolerance interval $[L, U]$ which
includes larger than $\gamma=0.953$ proportion of the distribution
of the output with probability $\beta=0.95$, then we should make
$100$ runs, see Table X. and select the lowest output as $L$ and
the largest as $U$. If $U$ is smaller than the technological limit
$U_{T}$, then the system is safe at the level $\gamma =0.953$,
$\beta=0.95$. This means that additional runs may produce an
output exceeding $U$ but this portion of runs is not larger than
$4.7$\% of the total number of runs. However, these rare output
values may be greater than the technological limit $U_{T}$.
Evidently, if $U$ is larger than $U_{T}$, the system must be
declared unsafe.

\vspace{0.4cm}

{\bf Table X.} {\footnotesize $\gamma$ values of tolerance
interval $[y(1),\;y(N)]$ for $\beta = 0.9, 0.95, 0.99$ and
$N=10(10)100(25)300$.}

\begin{center}
{\footnotesize
\begin{tabular}{|c|c|c|c|} \hline
{$N$} &
        \multicolumn{3}{c|}{$\gamma$ values} \\ \cline{2-4}
      & $\beta=0.90$ & $\beta=0.95$ & $\beta=0.99$ \\ \hline
    10&0.66315&0.60584&0.49565 \\
    20&0.81904&0.78389&0.71127 \\
    30&0.87643&0.85141&0.79845 \\
    40&0.90620&0.88682&0.84528 \\
    50&0.92443&0.90860&0.87448 \\
    60&0.93671&0.92336&0.89442 \\
    70&0.94557&0.93402&0.90890 \\
    80&0.95225&0.94207&0.91989 \\
    90&0.95747&0.94837&0.92851 \\
    100&0.96166&0.95344&0.93554 \\
    125&0.96924&0.96262&0.94813 \\
    150&0.97432&0.96877&0.95658 \\
    175&0.97796&0.97318&0.96268 \\
    200&0.98069&0.97650&0.96736 \\
    225&0.98282&0.97909&0.97087 \\
    250&0.98453&0.98118&0.97375 \\
    275&0.98593&0.98287&0.97618 \\
    300&0.98710&0.98429&0.97809 \\ \hline
    \end{tabular}}
    \end{center}

\vspace{0.3cm}

\begin{figure} [ht!]
\protect \centering{
\includegraphics[height=10cm, width=12cm]{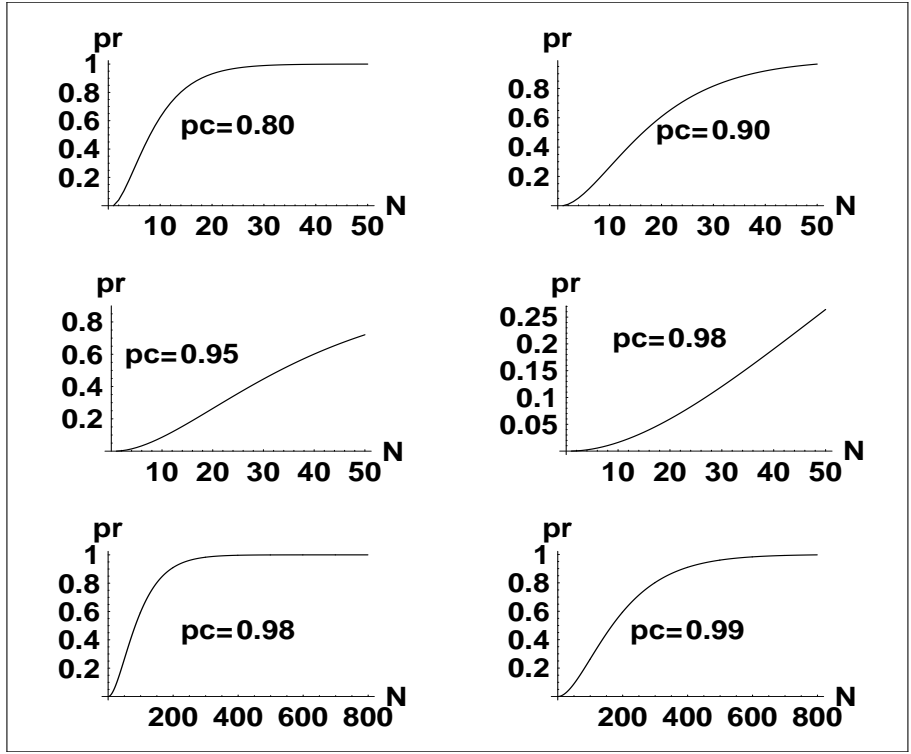}}\protect
\vskip 0.2cm \protect \caption{{\footnotesize The dependence of
the probability $\beta=pr$ on the the number of runs $N$ at
probability contents $\gamma=pc=0.8,\;0.9,\;0.95,\;0.98\;0.99$.}}
\label{fig8}
\end{figure}

To get some insight into relation (\ref{24}) we present the
probabilities $\beta$ versus $N$ for six $\gamma$ values, see Fig.
$8$. With increasing number of runs, each interpolated curve
reaches saturation, and $\beta$ tends to unity as $N$ tends to
infinity. The smaller is the $\gamma$ value the sooner comes the
saturation, because small $\gamma$ means that only a small
fraction of the calculated output is required to fall into the
given interval.

Small $\gamma$ value is not acceptable in safety analysis for
small $\gamma$ means that a large portion of output values may
fall outside the tolerance interval. Practically we need $\gamma
>0.95$. For example, if we wish the tolerance interval $[L,U]$ to
include larger than $\gamma=0.98$ proportion of the output values
with probability $\beta=0.95$, we need approximately $235$ runs in
order to get the proper $L$ and $U$. In spite of the large number
runs the probability content $\gamma=0.98$ is far from being
completely satisfactory. To achieve a better probability content,
say $\gamma=0.99$ with probability $\beta=0.95$, we need $473$
runs, which is practically hard to realize.

\subsubsection{Known cumulative distribution function}

Let us assume the cumulative distribution function $G(y)$ to be
known. However, one should emphasize that there are situations
where it would be particulary dangerous to make unwarranted
assumptions about the exact shape of distribution $G(y)$. In
general, the attempt to get an explicit expression for $\beta$ by
means of expression (\ref{20}) would fail. There is however one
exception, when $G(y)$ is of {\em normal distribution} $N(m,
\sigma)$ then exact formula can be obtained for
$\beta$.~\footnote{It is worth mentioning that if output variable
$y$ is a sum of a large number of small, statistically independent
random variable, then its distribution is almost normal. Now we
discuss the case when the output variable $y$ is of normal
distribution.}

We shall denote by $\tilde {y}_{N}$ the sample estimate of the
expectation value $m$ and by $\tilde {\sigma}_{N}^2$ that of the
variance $\sigma ^2$, i.e.
\begin{equation} \label{26}
\tilde{y}_{N} = \frac{1}{N} \sum_{k=1}^{N} y_{k}, \;\;\;\;\;\;
\mbox{and} \;\;\;\;\;\; \tilde{\sigma}_{N}^{2} = \frac{1}{N-1}
\sum_{k=1}^{N} (y_{k}-\tilde{y}_{N})^{2}.
\end{equation}
Let us construct two random variables, viz. \[ L = L(y_{1},
\ldots, y_{N};\lambda) = \tilde{y}_{N} -
\lambda\;\tilde{\sigma}_{N}\;\;\;\; \mbox{and} \;\;\;\; U =
U(y_{1}, \ldots, y_{N};\lambda) = \tilde{y}_{N} +
\lambda\;\tilde{\sigma}_{N},\] where the parameter $\lambda$
scales the length of the interval $[L,U]$. Denote by ${\mathcal
A}(\tilde {y}_{N}, \lambda \tilde{\sigma}_{N})$ the proportion of
the output distribution included between the limits $L(y_1,
\ldots, y_N; \lambda) = \tilde{y}_{N} - \lambda
\tilde{\sigma}_{N}$ and $U(y_1, \ldots, y_N; \lambda) =
\tilde{y}_{N} + \lambda \tilde{\sigma}_{N}$, i.e.
\begin{equation} \label{27}
{\mathcal A}(\tilde{y}_{N}, \lambda\tilde{\sigma}_{N}) =
\int_{L}^{U} g(y)\;dy = \frac{1}{\sqrt{2\pi}\sigma}\;\int_{L}^{U}
exp[-\frac{(y-m)^{2}}{2\sigma^{2}}]\;dy.
\end{equation}
Introducing new variable $z = (y-m)/\sigma$ we obtain
\begin{equation} \label{28}
{\mathcal A}(m+\sigma\tilde{z}_{N}, \lambda\tilde{\sigma}_{N}) =
\rho(\tilde{z}_{N}, \tilde{s}_{N}) =
\frac{1}{\sqrt{2\pi}}\;\int_{\ell_{N}}^{u_{N}} e^{-z^{2}/2}\;dz,
\end{equation}
where
\[ \tilde{z}_{N} = \frac{\tilde{y}_{N} - m}{\sigma} \;\;\;\;\;\; \mbox{and}
\;\;\;\;\;\; \tilde{s}_{N} = \frac{\tilde{\sigma}_{N}}{\sigma}, \]
while
\[ \ell_{N} = \tilde{z}_{N} - \lambda\;\tilde{s}_{N}
\;\;\;\;\;\; \mbox{and} \;\;\;\;\;\; u_{N} =  \tilde{z}_{N} +
\lambda \;\tilde{s}_{N}.  \] We stress again that $\rho (\tilde
{z}_{N},\tilde {s}_{N})$ is a random variable because in
expression (\ref{28}) the limits of the integral are random
variables.

\begin{theorem} \label{tr3}
For any given positive value of $\lambda$ the probability that
$\rho > \gamma$, where $0 << \gamma < 1$ is expressed by
\begin{equation} \label{29}
W(\lambda, \gamma, N) = 1 -
\sqrt{\frac{N}{2\pi}}\;\int_{-\infty}^{+\infty}K_{N-1}\left[(N-1)\;
\left(\frac{q(\mu, \gamma)}{\lambda}\right)^{2}\right]\;
e^{-N\mu^{2}/2}\;d\mu,
\end{equation}
where $K_{N-1}[\cdots]$ is the $\chi^{2}$ distribution with
$(N-1)$-degrees of freedom and $q(\mu, \gamma)$ is the solution of
the equation
\begin{equation} \label{30}
\frac{1}{\sqrt{2\pi}}\;\int_{\mu-q}^{\mu+q} e^{-x^{2}/2}\;dx =
\gamma.
\end{equation}
The value $\lambda$ determining the tolerance
interval~\footnote{If one-sided tolerance interval with upper
limit is needed, then Eq. (\ref{30}) should be replaced by \[
\frac{1}{\sqrt{2\pi}}\;\int_{-\infty}^{\mu+q} e^{-x^{2}/2}\;dx =
\gamma.\]} at a preassigned probability content $\gamma$ and a
preassigned significance level $\beta$ in the case of $N$ runs can
be calculated from the equation
\begin{equation} \label{31}
W(\lambda, \gamma, N) = \beta,
\end{equation}
and it is independent of unknown parameters $m$ and $\sigma$ of
the distribution function $G(y)$. The equation (\ref{31}) has
exactly one root in $\lambda$, since $W(\lambda, \gamma, N)$ is a
strictly increasing function of $\lambda$.
\end{theorem}

Proof of Theorem $3$ is given in Appendix III, since the
mathematical details are not relevant to the aim of the present
work. However, it is worth mentioning that an approximate
tolerance interval can be derived when $N$ is large (e.g. $N>50$).

\begin{theorem} \label{tr4}
The approximate two-sided tolerance interval is given by
\[ [\tilde{y}_{N} - \lambda_{a}(\gamma, \beta)\;\tilde{\sigma}_{N},
\;\;\tilde{y}_{N} + \lambda_{a}(\gamma,
\beta)\;\tilde{\sigma}_{N}],\] where
\begin{equation} \label{32}
\lambda_{a}(\gamma, \beta) = \sqrt{\frac{N-1}{Q_{N-1}(1-\beta)}}
q(1/\sqrt{N},\gamma).
\end{equation}
Here $Q_{N-1}(1-\beta)$ is {\em $(1-\beta)$-percentile} of the
$\chi^{2}$ distribution with $(N-1)$ degree of freedom and
$q(1/\sqrt{N},\gamma)$ is the root of the equation
\begin{equation} \label{33}
\frac{1}{\sqrt{2\pi}}
\int_{\frac{1}{\sqrt{N}}-q}^{\frac{1}{\sqrt{N}}+q}
e^{-z^{2}/2}\;dz = \gamma.
\end{equation}
{\footnotesize The $\lambda_{a}$ for the approximate one-sided
tolerance interval with upper limit can be calculated in the same
way, but Eq. (\ref{33}) has to be replaced by \[
\frac{1}{\sqrt{2\pi}} \int_{-\infty}^{\frac{1}{\sqrt{N}}+q}
e^{-z^{2}/2}\;dz = \gamma.\]}
\end{theorem}
Proof of Theorem 4 is given in Appendix IV.

\begin{center}
{\bf Table XI.} {\footnotesize $\lambda$ values of two-sided
tolerance intervals for the number of runs $N$=50(5)100}
\end{center}

\begin{center} {\footnotesize
\begin{tabular}{||c||c|c|c|c|c|c|c|c|c||} \hline \hline
&\multicolumn{3}{c|}{$\beta=0.90$}&\multicolumn{3}{c|}{$\beta=0.95$}&
\multicolumn{3}{c|}{$\beta=0.99$} \\ \hline $N\backslash \gamma$ &
0.90 & 0.95 & 0.99& 0.90 & 0.95 & 0.99& 0.90 & 0.95 & 0.99 \\
\hline
    50&1.916&2.284&3.001&1.996&2.379&3.126&2.162&2.576&3.385
    \\ \hline
    55&1.901&2.265&2.976&1.976&2.354&3.093&2.130&2.538&3.335
    \\ \hline
    60&1.887&2.248&2.956&1.958&2.333&3.066&2.103&2.506&3.293
    \\ \hline
    65&1.875&2.234&2.936&1.943&2.315&3.042&2.080&2.478&3.257
    \\ \hline
    70&1.865&2.222&2.920&1.929&2.299&3.021&2.060&2.454&3.225 \\
    \hline
    75&1.856&2.211&2.906&1.917&2.285&3.002&2.042&2.433&3.197 \\
    \hline
    80&1.848&2.202&2.894&1.907&2.272&2.986&2.026&2.414&3.173
    \\ \hline
    85&1.841&2.193&2.882&1.897&2.261&2.971&2.012&2.397&3.150
    \\ \hline
    90&1.834&2.185&2.872&1.889&2.251&2.958&1.999&2.382&3.130
    \\ \hline
    95&1.828&2.178&2.862&1.881&2.241&2.945&1.987&2.368&3.112
    \\ \hline
    100&1.822&2.172&2.854&1.874&2.233&2.934&1.977&2.355&3.096 \\
    \hline \hline
\end{tabular}}
\end{center}

\vspace{0.3cm}

\begin{figure} [ht!]
\protect \centering{
\includegraphics[height=6cm, width=9cm]{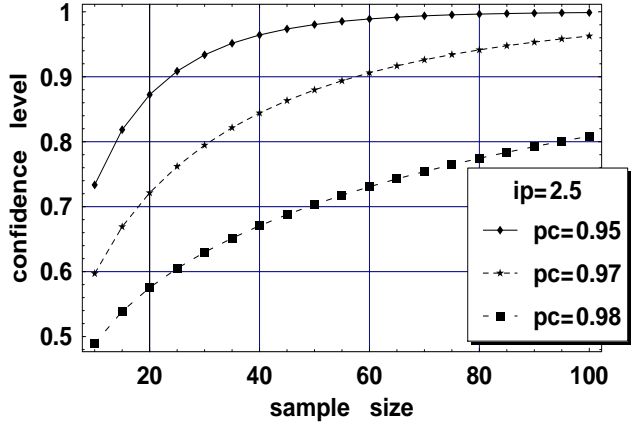}}\protect
\vskip 0.2cm \protect \caption{{\footnotesize Dependence of the
confidence level $\beta$ on the sample size $N$ at probability
contents $\gamma=pc=0.95, 0.97, 0.98$ when the interval parameter
$\lambda=ip=2.5$.}} \label{fig9}
\end{figure}

In order to give an impression of $\lambda$ values (i.e. of the
tolerance interval around the sample mean of the output variable),
Table XI. contains the $\lambda$ values~\footnote{More detailed
tables can be found in \cite{odeh80}.} associated with often used
$\gamma$ and $\beta$ for the sample sizes $N$=50(5)100. One can
see that at $N$=100 the tolerance interval which includes $95${\%}
of the distribution with $95${\%} probability is given by
\[ \left[\tilde{y}_{100} - 2.23 \tilde{\sigma}_{100},\;
\tilde{y}_{100} + 2.23 \tilde {\sigma}_{100} \right].\] If that
interval~\footnote{If one-sided tolerance interval with upper
limit is needed, then $\lambda=2.23$ has to be replaced by
$\lambda=1.75$!} lies within $[L_{T}, U_{T}]$ then the system is
safe on level $\gamma=0.95$ and $\beta=0.95$.

\begin{figure} [ht!]
\protect \centering{
\includegraphics[height=6cm, width=9cm]{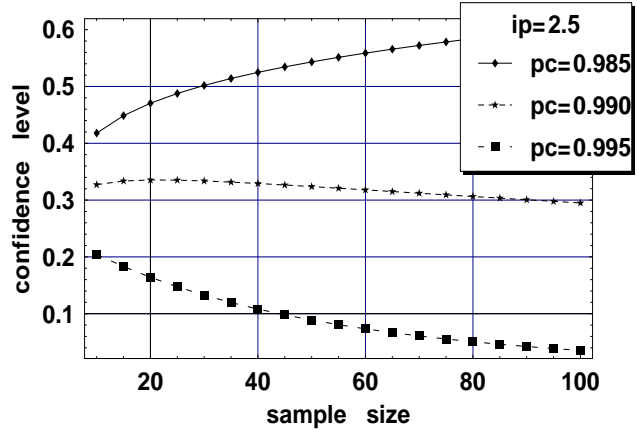}}\protect
\vskip 0.2cm \protect \caption{{\footnotesize Dependence of the
confidence level $\beta$ on the sample size $N$ at probability
contents larger than the critical value $\gamma_{crt}\approx
0.98758$ and provided the interval parameter $\lambda=ip=2.5$ is
fixed.}} \label{fig10}
\end{figure}

Fig. $9$ shows convincingly the interrelations between the basic
characteristics of the tolerance intervals for a normal
distribution.  As expected the confidence level $\beta$ increases
with increasing sample size  $N$ provided that the coverage
$pc=\gamma $ and the interval parameter $ip=\lambda$ are fixed.

However, if the fixed coverage $\gamma$ exceeds a critical value
$\gamma_{crt}\approx 0.98758$ when $\lambda=ip=2.5$, then  one can
observe an "anomalous" behavior of the dependence $\beta$ on $N$,
as shown in Fig. $10$. It is seen that the probability $\beta$ of
finding the proportion $\gamma > \gamma_{crt}$ of the distribution
$G(y)$ in the interval $(\tilde{z}_{N} - \lambda\;\tilde{s}_{N},
\;\tilde{z}_{N} + \lambda\;\tilde{s}_{N})$ decreases with
increasing sample size $N > N_{crt}$, where $N_{crt}$ depends on
both $\lambda$ and $\gamma$. The explanation is straightforward:

\begin{figure} [ht!]
\protect \centering{
\includegraphics[height=6cm, width=9cm]{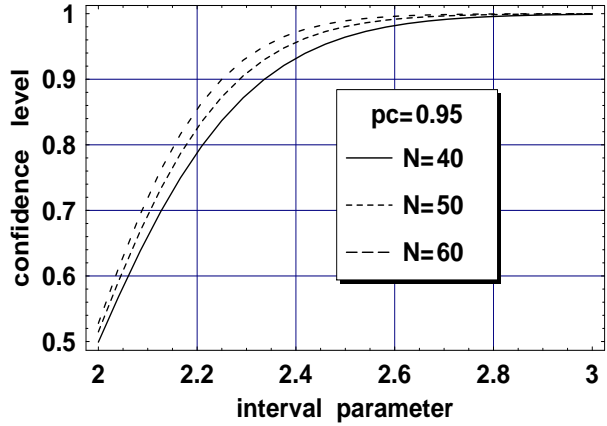}}\protect
\vskip 0.2cm \protect \caption{{\footnotesize Dependence of the
confidence level $\beta$ on the interval parameter $\lambda=ip$ at
probability content $\gamma=pc=0.95$ for three sample sizes
$N=40,\;50,\;60$.}} \label{fig11}
\end{figure}

\begin{figure} [ht!]
\protect \centering{
\includegraphics[height=6cm, width=9cm]{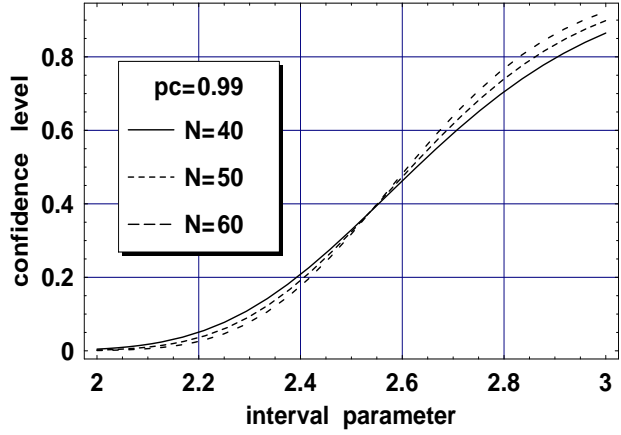}}\protect
\vskip 0.2cm \protect \caption{{\footnotesize Dependence of the
confidence level $\beta$ on the interval parameter $\lambda=ip$ at
probability content $\gamma=pc=0.99$ higher than the critical
value for three sample sizes $N=40,\;50,\;60$.}} \label{fig12}
\end{figure}
\noindent since
\[ \lim_{N \rightarrow \infty} \tilde{z}_{N}
\stackrel{p}{=} 0\;\;\;\;\;\; \mbox{and} \;\;\;\;\;\;
\lim_{N\rightarrow \infty} \tilde{s}_{N} \stackrel{p}{=} 1, \] it
is evident that
\[ \lim_{N \rightarrow \infty}\rho(\tilde{z}_{N}, \tilde{s}_{N})
\stackrel{p}{=} \gamma_{crt}, \;\;\;\;\;\; \mbox{where}
\;\;\;\;\;\;  \gamma_{crt} =
\frac{1}{\sqrt{2\pi}}\;\int_{-\lambda}^{+\lambda}
e^{-x^{2}/2}\;dx,  \]  consequently, if $\gamma = \gamma_{crt} +
\delta$, where $0 < \delta < 1 - \gamma_{crt}$, then
\[ \lim_{N \rightarrow \infty} {\mathcal P}\{\vert \rho(\tilde{z}_{N},
\tilde{s}_{N}) - \gamma_{crt}\vert > \delta\} = 0, \] i.e.
$\;\;{\mathcal P}\{\vert \rho(\tilde{z}_{N}, \tilde{s}_{N}) -
\gamma_{crt}\vert > \delta\}\;\;$ is a monotonously decreasing
function of $N > N_{crt}$. It easy to show
that~\footnote{Introducing the notations:
\[ \{\rho(\tilde{z}_{N}, \tilde{s}_{N}) \leq \gamma_{crt}-\delta\}
= {\mathcal A}_{N}^{(-)}\] and
\[ \{\rho(\tilde{z}_{N}, \tilde{s}_{N}) > \gamma_{crt}+\delta\}
= {\mathcal A}_{N}^{(+)}, \] and taking into account that
${\mathcal A}_{N}^{(+)} \cap {\mathcal A}_{N}^{(-)} = \emptyset$,
we can write that
\[ {\mathcal P}\{\vert\rho(\tilde{z}_{N}, \tilde{s}_{N}) -
\gamma_{crt}\vert > \delta\} = {\mathcal P}\{{\mathcal
A}_{N}^{(+)} \cup  {\mathcal A}_{N}^{(-)}\} > {\mathcal
P}\{\rho(\tilde{z}_{N}, \tilde{s}_{N}) > \gamma_{crt}+\delta\}.
\]} \[ {\mathcal P}\{\vert \rho(\tilde{z}_{N}, \tilde{s}_{N}) -
\gamma_{crt}\vert > \delta\} > {\mathcal P}\{\rho(\tilde{z}_{N},
\tilde{s}_{N}) > \gamma_{crt}+\delta\} = \beta, \] and so one can
state that $\beta$ decreases with increasing $N > N_{crt}$ if
$\gamma > \gamma_{crt}$ provided $\lambda$ is fixed.

It is not superfluous to know how does the confidence level
$\beta$ depend on the interval parameter $\lambda$ at a fixed
probability content (coverage) $\gamma$ and at a given sample size
$N$. Fig. $11$ shows this dependence at $\gamma=pc=0.95$ for three
sample sizes $N=40, 50, 60$. What we see completely corresponds to
our expectations, however, as seen in Fig. $12$, the character of
$\beta$ vs. $\lambda$ curves is radically changing. The
explanation is the same as in the case of Fig. $10$.

\section{Several output variables}

Now we assume the output to comprise $n$ variables. Let these
variables be $y_{1}, \ldots, y_{n}$. If they are statistically
completely independent~\footnote{There are many fairly good
statistical tests to prove the independence of random variables.}
we can apply the results  of previous Sections, otherwise we need
new considerations. Let $G(y_{1}, \ldots, y_{n})$ be the unknown
joint cumulative distribution function of the output variables,
furthermore, let
\begin{equation} \label{34}
\underline{{\mathcal S}}_{N} = \left(\begin{array}{cccc}
y_{11} & y_{12} & \ldots & y_{1N} \\
y_{21} & y_{22} & \ldots & y_{2N} \\
\vdots & \vdots & \ddots & \vdots \\
y_{n1} & y_{n2} & \ldots & y_{nN}
\end{array} \right)
\end{equation}
be the sample matrix obtained in $N>>2n$ independent observations
(runs). Introducing the $n$-components vector
\[ \vec{y}_{k} = \left( \begin{array}{c} y_{1k} \\ y_{2k} \\
\vdots \\ y_{nk} \end{array} \right),\] the sample matrix can be
written in the form:
\[ \underline{{\mathcal S}}_{N} = \left\{\vec{y}_{1}, \ldots,
\vec{y}_{N}\right\}. \] By using proper statistical methods for
testing the sample matrix we can make useful probabilistic
statement about the safety of system operation.

First, we will show how to generalize the method of {\em sign test
for several output variables}, and then we will deal with the
problem of setting {\em tolerance limits for more than one random
variable}.

\subsection{Sign test}

For the sake of simplicity we are going to deal with two output
variables $y_{1}$ and $y_{2}$ provided their joint distribution
function $G(y_{1}, y_{2})$  is unknown, but continuous at least
from right (or from left) in both variables. Let us accept that
the system operation can be declared safe if the requirement
$\{y_{1} < U_{T}^{(1)}, \; y_{2} < U_{T}^{(2)}\}$ is realized with
probability
\begin{equation} \label{35}
p_{12} = {\mathcal P}\{y_{1} < U_{T}^{(1)}, \; y_{2} <
U_{T}^{(2)}\}
\end{equation}
near the unity. Here $U_{T}^{(1)}$ and $U_{T}^{(2)}$ are the limit
values defined by technology, and they define the {\em acceptance
region} of the $(y_{1}, y_{2})$ plane.  Since the $p_{12}$ is
unknown, the task is to construct from the sample a confidence
interval $[\gamma_{L}^{(1,2)},\; \gamma_{U}^{(1,2)}]$ which covers
the $p_{12}$ with probability $\beta_{12}$. In most of the cases
it is sufficient to calculate the $\gamma_{L}^{(1,2)}$ only and to
use the interval $[\gamma_{L}^{(1,2)},\;1]$ as confidence
interval. Let the 2-components vectors
\[ \vec{y}_{k} = \binom{y_{1k}}{y_{2k}},  \;\;\;\;\;\; k=1, \ldots, N \]
be elements of a sample $\underline{{\mathcal S}}_{N}$ obtained by
$N$ independent observations. One should emphasize that
$\vec{y}_{j}$ and $\vec{y}_{k}$ are independent if $j \neq k$, but
the components of a given sample vector are not.

In order to use a terminology as simple as possible, the event
$\{y_{1} < U_{T}^{(1)}, \; y_{2} < U_{T}^{(2)}\}$ will be called
success. Define now the function
\[
\Delta\left(U_{T}^{(1)}-y_{1k}\right)\;\Delta\left(U_{T}^{(2)}-y_{2k}\right)
= \left\{ \begin{array}{ll} 1, & \mbox{if $y_{1k} < U_{T}^{(1)}$
and $y_{2k} < U_{T}^{(2)}$, } \\
\mbox{} & \mbox{} \\
0, & \mbox{otherweise}, \end{array} \right. \] and introduce the
statistical function
\begin{equation} \label{36}
z_{N}^{(1,2)} = \sum_{k=1}^{N}
\Delta\left(U_{T}^{(1)}-y_{1k}\right)\;\Delta\left(U_{T}^{(2)}-y_{2k}\right),
\end{equation}
which gives the number of successes in a sample of size $N$. Since
$z_{N}^{(1,2)}$ is the sum of $N$ independent random variables
with values either $1$ or $0$, it is obvious that $z_{N}^{(1,2)}$
is of binomial distribution. By using the notation
\[ {\mathcal P}\{\Delta\left(U_{T}^{(1)}-y_{1}\right)\;
\Delta\left(U_{T}^{(2)}-y_{2}\right)=1\} = \]
\[ = {\mathcal P}\{y_{1} < U_{T}^{(1)}, \; y_{2} < U_{T}^{(2)}\}
= p_{12},  \] we can write
\[ {\mathcal P} \{z_{N}^{(1,2)}=k \} = \binom{N}{k} p_{12}^{k}\;(1-p_{12})^{N-k},
\;\;\;\;\;\; \forall \;\; k=0,1, \ldots, N,\] and this is the
point where we can use from the results of Subsection $3.3$.

Now, we would like to make a trivial but important amendment.
Define two statistical functions:
\[ z_{N}^{(1)} = \sum_{i=1}^{N} \Delta\left(U_{T}^{(1)} -
y_{1i}\right) \;\;\;\; \mbox{and} \;\;\;\; z_{N}^{(2)} =
\sum_{j=1}^{N} \Delta\left(U_{T}^{(2)} - y_{2j}\right). \]
Clearly, $z_{N}^{(1)}$ and $z_{N}^{(2)}$ are not independent, but
both of them are sum of $N$ independent random variables with
values either $1$ or $0$, consequently one can write
\[ {\mathcal P}\{z_{N}^{(1)}=i\} = \binom{N}{i} p_{1}^{i}
(1-p_{1})^{N-i} \] and
\[ {\mathcal P}\{z_{N}^{(2)}=j\} = \binom{N}{j} p_{2}^{j}
(1-p_{2})^{N-j},  \]
\[ i,j = 1, \ldots, N, \]
where
\[ p_{\ell} = {\mathcal P}\{y_{\ell} < U_{T}^{(\ell)}\} =
{\mathcal P}\{\Delta\left(U_{T}^{(\ell)} - y_{\ell}\right) = 1\},
\] \[ \ell = 1, 2, \]
are {\em unknown probabilities}. By using the samples ${\mathcal
S}_{N}^{(1)} = \{y_{1i}, \;\; i=1, \ldots, N\}$ and ${\mathcal
S}_{N}^{(2)} = \{y_{2j}, \;\; j=1, \ldots, N\}$ separately with
help of the method described in Subsection $3.3$ we can construct
two random intervals $[\gamma_{L}^{(1)}, \;1]$ and
$[\gamma_{L}^{(2)}, \;1]$ covering $p_{1}$ as well as $p_{2}$ with
probabilities $\beta_{1}$ and $\beta_{2}$, respectively. Obviously
it could be occurred that the levels $(\beta_{1}\vert
\gamma_{L}^{(1)})$ and $(\beta_{2}\vert \gamma_{L}^{(2)})$ support
the statement that the samples  ${\mathcal S}_{N}^{(1)}$ and
${\mathcal S}_{N}^{(2)}$ {\em separately} do not contradict to the
requirement of safe operation, however, from this one cannot
conclude that the operation of the system is safe on a preassigned
level for variables $y_{1}$ and $y_{2}$ tested {\em jointly}. The
reason is clear: the output variables $y_{1}$ and $y_{2}$ are not
independent, and in this case we have to know weather the value
$p_{12} = {\mathcal P}\{y_{1} < U_{T}^{(1)}, y_{2} <
U_{T}^{(2)}\}$ is covered by the interval
$[\gamma_{L}^{(1,2)},\;1]$ with a preassigned probability
$\beta_{12}$. Clearly, $\gamma_{L}^{(1,2)} \leq
\min\{\gamma_{L}^{(1)}, \gamma_{L}^{(2)}\}$, therefore
$\gamma_{L}^{(1)}$ and $\gamma_{L}^{(2)}$ do not contain
sufficient information to declare that the operation of the system
is safe. The procedure should be as follows: firstly test the
hypothesis that the output variables $y_{1}$ and $y_{2}$ are
dependent, and if this is the case, estimate the probability of
the event $\{y_{1} < U_{T}^{(1)}, y_{2} < U_{T}^{(2)}\}$, and not
the events $\{y_{1} < U_{T}^{(1)}\}$ and $\{y_{2} < U_{T}^{(2)}\}$
separately.

Finally, we would like to note that the {\em generalization of the
sign test for $n>2$ output variables} is straightforward: we have
to use the statistical function
\[ z_{N}^{(1,\ldots,n)} = \sum_{k=1}^{N}\;\prod_{j=1}^{n}
\Delta\left(U_{T}^{(j)}-y_{jk}\right), \] in order to obtain the
sum of $N$ independent random variables, and then the further
steps will be the same as they were in Subsection $3.3$.

\subsubsection{Illustration}

Now we want to present an example to show how the sign test method
is working. By using Monte Carlo simulation we have generated two
samples {\bf a} and {\bf b}. Both are consisting of $N=100$ value
pairs due to the population of a bivariate normal distribution
with parameters $m_{1}=m_{2}=0$ and $\sigma_{1}=\sigma_{2}=1$, but
the correlation coefficient is $C=0.1$ in {\bf a}, while $C=0.7$
in {\bf b}.

\begin{figure} [ht!]
\protect \centering{
\includegraphics[height=14cm, width=10cm]{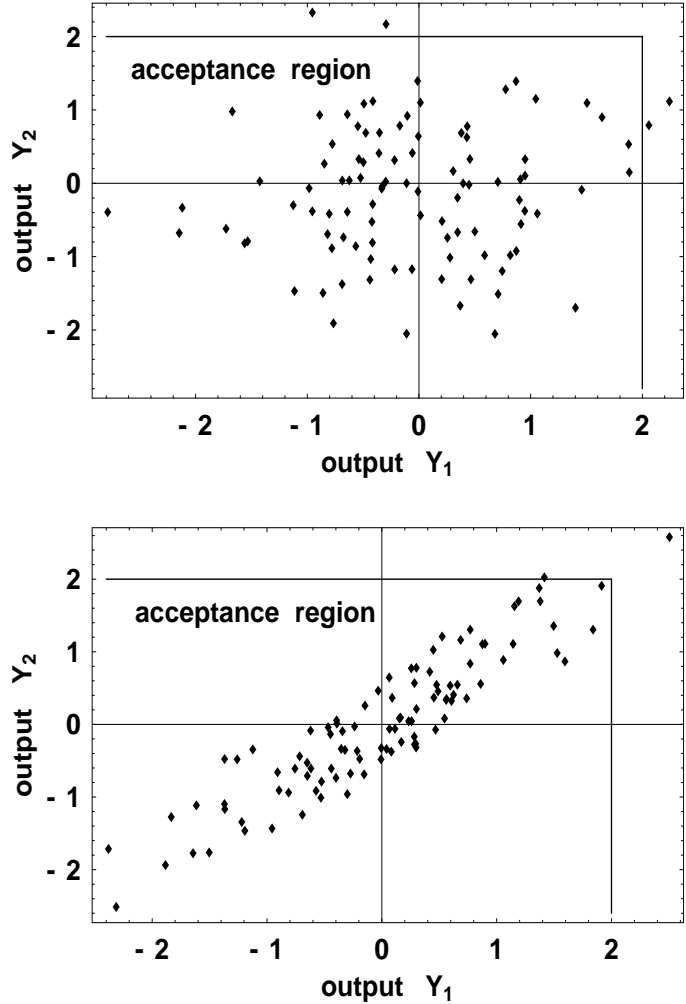}}\protect
\vskip 0.2cm \protect \caption{{\footnotesize Sample vectors
denoted by points in the sample plane $(y_{1}, y_{2})$ and the
acceptance region defined by technological requirements. Upper
figure (sample {\bf a}) and lower figure (sample {\bf b}) refer to
correlation coefficients $C=0.1$ and $C=0.7$, respectively. }}
\label{fig13}
\end{figure}

One can see in Fig. $13$ that in the sample {\bf a} four, while in
the {\bf b} two observations out of $N=100$ can be found in
rejection region.

From Table IX. one can read that in the case of sample {\bf a} the
interval $[0.9108, 1]$ covers the parameter $p_{12}$ with
probability $\beta_{12}=0.95$, at the same time both $p_{1}$ and
$p_{2}$ are covered  by the interval $[0.9383, 1]$ with
$\beta_{1}=\beta_{2}=0.95$. The level $(0.9383\vert 0.95)$ is not
"very good", but better than $(0.9108\vert 0.95)$, however, in the
decision about the safety one should take evidently into account
the level calculated for the parameter $p_{12}$, and not those
calculated separately for $p_{1}$ and $p_{2}$.

Testing the sample {\bf b} which shows a strong correlation
between the variables $y_{1}$ and $y_{2}$, we find that the
confidence interval $[0.9383, 1]$ covers the parameter $p_{12}$
with probability $\beta_{12}=0.95$. Consequently, we can state
with probability $0.95$ that the chance of the event $\{y_{1} <
U_{T}^{(1)}, y_{2} < U_{T}^{(2)}\}$ is higher than the value
$0.9383$, i.e. we are able to declare that the operation of the
system is safe on the level $(0.95\vert 0.9383)$ only.  The
parameters $p_{1}$ and $p_{2}$ are covered by intervals $[0.9534,
1]$ and $[0.9383, 1]$, respectively, with the prescribed
probability $\beta_{1}=\beta_{2}=0.95$, however, these values are
not informative for the safety of the system.

This simple example shows convincingly that the tests performed
separately on output variables which are depending on one another
could bring about false decision concerning the safety of the
system operation.

\subsection{Tolerance region}

The problem of setting tolerance limits for output variables
$y_{1}, \ldots, y_{n}$ can be formulated as follows. Assume that
the unknown joint distribution function $G(y_{1}, \ldots, y_{n})$
is absolute continuous, i.e. it has a joint density function
$g(y_{1}, \ldots, y_{n})$. For some given positive values $\gamma
< 1$ and $\beta < 1$ we have to construct $n$ pairs of random
variables $L_{j}(y_{1}, \ldots, y_{n})$ and $U_{j}(y_{1}, \ldots,
y_{n})\;\; j=1,\ldots,n$ such that the probability that
\begin{equation} \label{37}
\int_{L_{1}}^{U_{1}} \cdots \int_{L_{n}}^{U_{n}} g(y_{1}, \ldots,
y_{n})\;dy_{1} \cdots dy_{n} > \gamma,
\end{equation}
holds is equal to $\beta$. A natural extension of the procedure
applied previously to the one variable case would seem the right
selection. Unfortunately that choice does not provide the required
solution since the probability of the inequality (\ref{37})
depends on the unknown joint density function $g(y_{1}, \ldots,
y_{n})$. Our task is to find a reasonable procedure such that the
probability $\beta$ is independent of $g(y_{1}, \ldots, y_{n})$.
It can be shown that such a procedure exists but its uniqueness
has not been proven yet.

Since the distribution function $G(y_{1}, \ldots, y_{n})$ is
continuous, we can state that no two elements of the sample matrix
$\underline{{\mathcal S}}_{N}$ are equal. The sequence of rows in
the sample matrix $\underline{{\mathcal S}}_{N}$ can be arbitrary,
reflecting the fact that we number the output variables
arbitrarily.

Let us choose the first row of the sample matrix, and arrange its
elements in order of increasing magnitude $y_{1}(1), y_{1}(2),
\ldots, y_{1}(N)$. Select from these $y_{1}(r_{1})$ as $L_{1}$ and
$y_{1}(s_{1}) > y_{1}(r_{1})$ as $U_{1}$. Let $i_1, i_2, \ldots,
i_{s_1-r_1-1}$ stand for the original column indices of elements
$y_{1}(r_{1}+1), y_{1}(r_{1}+2), \ldots, y_{1}(s_{1}-1)$. In the
next step, choose the second row, the $N$ observed values of the
output variable $y_{2}$ and arrange the part $y_{2i_1}, y_{2i_2},
\ldots, y_{2i_{s_1-r_1-1}}$ of its elements in increasing order to
obtain $y_{2}(1)< y_{2}(2) < \cdots < y_{2}(s_{1}-r_{1}-1)$. From
among these, $y_{2}(r_{2})$ and $y_{2}(s_{2})>y_{2}(r_{2})$ are
selected for $L_{2}$ and $U_{2}$ and evidently $r_2 \ge r_1,\;\;
s_2 \le s_1-r_1-1$. We continue this imbedding procedure to the
last row of the sample matrix and define a $n$-dimensional
volume~\footnote{This $n$-dimensional volume is the tolerance
region which is nothing else than a subspace of an n-dimensional
Euclidian space.}
\[ {\mathcal V}_{n} =\{[L_{1}, U_{1}]\times[L_{2}, U_{2}]\times
\cdots \times [L_{n}, U_{n}]\}, \] where
\[ L_{j} = y_{j}(r_{j}), \;\;\;\;\;\;  U_{j} = y_{j}(s_{j}), \]
and
\[ r_{j} \geq r_{j-1} \geq \cdots \geq r_{1},\] while
\[r_{j} < s_{j} \leq s_{j-1}-r_{j-1}-1, \;\;\;\;\;\; \forall\;\;
j=2, \ldots, n.\]

\begin{theorem} \label{tr5}
In the case of $n \geq 2$ dependent output variables with
continuous joint distribution function $G(y_{1}, \ldots, y_{n})$
it is possible to construct $n$-pairs of random intervals
$[L_{j},\;U_{j}], \;\; j=1,\ldots, n$ such that the probability of
the inequality
\[ \int_{L_{1}}^{U_{1}} \cdots \int_{L_{n}}^{U_{n}} g(y_{1}, \ldots,
y_{n})\;dy_{1} \cdots dy_{n} > \gamma, \] is free of $g(y_{1},
\ldots, y_{n})$ and is given by
\[ {\mathcal P}\left\{\int_{L_{1}}^{U_{1}} \cdots
\int_{L_{n}}^{U_{n}} g(y_{1}, \ldots, y_{n})\;dy_{1} \cdots dy_{n}
> \gamma \right\} =  \]
\begin{equation} \label{38}
= 1 - I\left(\gamma, s_{n}-r_{n}, N-s_{n}+r_{n}+1\right) = \beta.
\end{equation}
Here function $I(\cdots)$ is the regularized incomplete
beta-function and
\begin{equation} \label{39}
s_{n} \leq s_{n-1} - r_{n-1} - 1 \leq s_{1} -
\sum_{j=1}^{n-1}(r_{j} + 1) \;\;\;\; \mbox{and} \;\;\;\; r_{n}
\geq r_{n-1} \geq \cdots \geq r_{1}.
\end{equation}
\end{theorem}

\vspace{0.3cm}

Proof of Theorem 5 is given in Appendix V.

\subsubsection{Illustrations}

In several practical applications the choice
$r_{1}=r_{2}=\cdots=r_{n}=1$ and $s_n=N-2(n - 1)$ can be advised,
hence the confidence level $\beta$ for a {\em two-sided tolerance
region} is given by
\begin{equation} \label{40}
\beta = 1 - I\left(\gamma, N-2n+1, 2n\right) =
\sum_{j=0}^{N-2n}\binom{N}{j}\;\gamma^{j}\;(1-\gamma)^{N-j}.
\end{equation}
The structure of expression (\ref{40}) is remarkably similar to
that of expression (\ref{22}), which refers to the one output
variable case. Furthermore, if the lower limits $L_{j}=-\infty,
\;\; \forall \; j=1,\ldots,n$, i.e if \[ r_{1}=r_{2}= \cdots =
r_{n} = 0 \;\;\;\; \mbox{and} \;\;\;\; s_{n} = N - n + 1,\]  then
one obtains the confidence level
\begin{equation} \label{41}
\beta = 1 - I\left(\gamma, N-n+1, n\right) =
\sum_{j=0}^{N-n}\binom{N}{j}\;\gamma^{j}\;(1-\gamma)^{N-j}
\end{equation}
for {\em one-sided tolerance region}.

In many practical cases it is sufficient to use {\em one-sided
tolerance regions} (limited from above). If $n=2$, i.e. if two
mutually dependent output variables $y_{1}$ and $y_{2}$ are
tested, then from (\ref{41}) one obtains
\begin{equation} \label{42}
\beta = 1 - \gamma^{N} - N(1-\gamma) \gamma^{N-1}
\end{equation}
which is exactly the same as (\ref{24}) derived for the two-sided
tolerance interval for one output variable. Here, it is worthwhile
to cite two sentences from \cite{makai03}. {\em "There are several
ways to interpret even a simple mathematical formula. The problem
under consideration decides which interpretation we need.
Notwithstanding, we should carefully prove the appropriateness of
the interpretation chosen."}

Perhaps, it is not superfluous to show how to determine the
two-\-dimen\-sional one-sided tolerance region for output
variables $y_{1}$ and $y_{2}$. First, calculate from (\ref{42})
the number of observations $N$ needed for the preassigned safety
level $(\beta\vert \gamma)$. Secondly, create the the sample
\[ \underline{{\mathcal S}}_{N} = \left(\begin{array}{ll}
y_{11}, y_{12}, \ldots, y_{1N} & \\
y_{21}, y_{22}, \ldots, y_{2N} \end{array} \right), \] and arrange
the elements of the first row in increasing order. We obtain the
matrix
\[ \left(\begin{array}{cccc}
y_{1}(1),& y_{1}(2), & \ldots, & y_{1}(N)  \\
y_{2i_{1}}, & y_{2i_{2}}, & \ldots, & y_{2i_{N}} \end{array}
\right),\] and choose the element $y_{1}(N)$ as upper limit for
$y_{1}$, i.e. $U_{1}=y_{1}(N)$. Thirdly, search the largest
element in the series $y_{2i_{1}}, y_{2i_{2}}, \ldots,
y_{2i_{N-1}}$ that gives the upper limit for $y_{2}$, i.e.
$U_{2}=\max_{1\leq j \leq N-1} y_{2i_{j}}$, and finally, construct
the region $[-\infty, U_{1}]\times[-\infty, U_{2}]$ which is
tolerance region of variables $(y_{1}$ and $y_{2})$. Clearly, if
$U_{1}<U_{T}^{(1)}$ and $U_{2}<U_{T}^{(2)}$, then we can state
that the operation of the system is safe on the level $(\beta\vert
\gamma)$ for the jointly tested two variables $(y_{1}$ and
$y_{2})$.

In order to compare the number of runs needed to determine {\em
two-sided tolerance regions} at a given ($\beta\vert \gamma$)
level for $n=1,2$, and $3$ mutually dependent output variables
with unknown distributions, we compiled Table XII. In order to
achieve the usual safety level $(0.95\vert 0.95)$ we need $N=153$
observations in the case of two and $N=207$ observations in the
case of three output variables. The number of observations (runs)
needed to meet stringent requirement, e.g. with three output
variables the level $(\gamma=0.98\vert \beta=0.98)$ we need
$N=598$ runs. Therefore it seems to be inevitable to seek methods
with lower computational demands.

\vspace{0.3cm}

\begin{center}
{\bf Table XII.} {\footnotesize Number of runs needed to determine
the two-sided tolerance region for $n=1,2,3$ output variables at
listed $\gamma, \beta$ values.}
\end{center}

\begin{center}
{\footnotesize
\begin{tabular}{|c|c|c|c|c|c|c|} \hline
$\beta \backslash \gamma$ & 0.95 & 0.96 & 0.97 & 0.98 & 0.99 & $n$  \\
\hline   &  93 & 117 & 156 & 235 & 473  & 1 \\
 0.95    & 153 & 191 & 256 & 385 & 773  & 2 \\
         & 207 & 260 & 348 & 523 & 1049 & 3 \\ \hline
         &  98 & 123 & 165 & 249 & 499  & 1 \\
 0.96    & 159 & 200 & 267 & 402 & 806  & 2 \\
         & 215 & 269 & 360 & 542 & 1086 & 3 \\ \hline
         & 105 & 132 & 176 & 266 & 533  & 1 \\
 0.97    & 167 & 210 & 281 & 422 & 848  & 2 \\
         & 224 & 281 & 376 & 565 & 1134 & 3 \\ \hline
         & 114 & 143 & 192 & 289 & 581  & 1 \\
 0.98    & 179 & 224 & 300 & 451 & 905  & 2 \\
         & 237 & 297 & 397 & 598 & 1199 & 3 \\ \hline
         & 130 & 163 & 218 & 329 & 661  & 1 \\
 0.99    & 197 & 248 & 331 & 499 & 1001 & 2 \\
         & 258 & 324 & 433 & 651 & 1307 & 3 \\ \hline
\end{tabular}}
\end{center}

\vspace{0.3cm}

In order to provide some insight, let us consider the following
example. We have two output variables $y_{1}$ and $y_{2}$, their
the joint distribution function is known:
\begin{equation} \label{43}
g(y_1, y_2) = \frac{1}{2\pi \sqrt{1 - C^2}}\;\exp
\left[-\frac{1}{2(1-C^2)}\;(y_1^2 - 2Cy_1 y_2 + y_2^2)\right],
\end{equation}
where $\left|C\right| \le 1$ is the correlation coefficient of
variables $y_{1}$ and $y_{2}$. We are interested in the
relationship between the significance level $\beta$ and
probability content of a given two dimensional region
$[L,U]=[L_{1}, U_{1}]\times [L_{2}, U_{2}]$ at the number of runs
$N=50(50)200$.

\vspace{0.3cm}

\begin{center}
{\bf Table XIII.} {\footnotesize Levels of significance $\beta$ of
two-sided tolerance regions for two output variables at listed
$\gamma$ and $N$ values.}
\end{center}

\begin{center}
{\footnotesize
\begin{tabular}{|c|c|c|c|c|c|} \hline
$N \backslash \gamma$ & 0.95 & 0.96 & 0.97 & 0.98 &   \\
\hline   &  0.8831 & 0.7547 & 0.5351 & 0.2376 &  $C=0.1$ \\
 50      &  0.9433 & 0.8775 & 0.7442 & 0.4970 &  $C=0.9$ \\
         &  0.2396 & 0.1391 & 0.0628 & 0.0178 &  DF    \\ \hline
         &  0.9109 & 0.8836 & 0.6297 & 0.2121 & $C=0.1$ \\
 100     &  0.9911 & 0.9590 & 0.8488 & 0.4970 &  $C=0.9$ \\
         &  0.7422 & 0.5705 & 0.3528 & 0.1410 &  DF    \\ \hline
         &  0.9933 & 0.9443 & 0.6871 & 0.1894 &  $C=0.1$ \\
 150     &  0.9981 & 0.9869 & 0.9044 & 0.5554 & $C=0.9$ \\
         &  0.9452 & 0.8542 & 0.6616 & 0.3528 &  DF    \\ \hline
         &  0.9986 & 0.9683 & 0.7380 & 0.1612 &  $C=0.1$ \\
 200     &  0.9998 & 0.9955 & 0.9414 & 0.5779 &  $C=0.9$ \\
         &  0.9910 & 0.9605 & 0.8528 & 0.5685 &  DF    \\ \hline
\end{tabular}}
\end{center}

\vspace{0.3cm}

\begin{figure} [ht!]
\protect \centering{
\includegraphics[height=6cm, width=9cm]{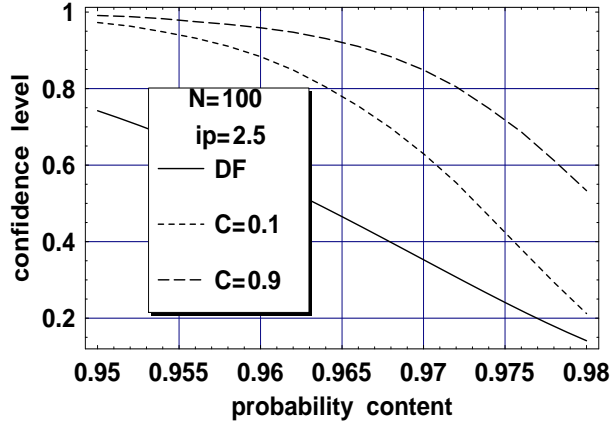}}\protect
\vskip 0.2cm \protect \caption{{\footnotesize Two output
variables.}} \label{fig14}
\end{figure}

Now we can proceed in two ways. The first way is to fix the
fraction of the samples to fall into the given interval $[L, U]$,
and to determine the associated probability $\beta$, from Eq.
(\ref{41}), these numbers are in row DF (referring to {\em
Distribution Free}). The second way is to use the known joint
distribution function, calculate the estimates of variances
$\tilde{\sigma}_i$ for $i=1,2$ from $N$ runs and define the
interval $[L_{i}, U_{i}]=[-2.5\tilde{\sigma}_i,
+2.5\tilde{\sigma}_i]$. From $10^{5}$ random cases we estimated
the $\beta$ value, see Table XIV. These values are given for two
correlation coefficient in rows $C=0.1$ and $C=0.9$.

As we see in Table XIII. the order statistics gives lower $\beta$
values, in most of the cases, compared to those obtained by using
the density function $g(y_{1}, y_{2}, C)$.~\footnote{Without going
into details, we mention only there exists a critical value
$\gamma_{cr}(C)$ such that for $\gamma > \gamma_{cr}(C)$ for all
$N_{1} < N_{2}$ we have $\beta(\gamma, N_{1}) > \beta(\gamma,
N_{2})$. The critical value $\gamma_{cr}(C)$ is defined by the
integral
$\int_{-\lambda}^{\lambda}\;\int_{-\lambda}^{\lambda}g(y_{1},
y_{2}, C)\;dy_{1}\;dy_{2}$, similarly to that proved in
one-dimensional case. See sub-subsection 3.4.2! The decrease of
$\beta$ with increasing $N$ can be so large for $N>100$ that the
$\beta$ becomes smaller than the value obtained from the order
statistics. This is the case for some $\beta$ values with $C=0.1$
in Table XIII. when $N>100$ and $\gamma > \gamma_{cr}(0.1) =
0.9753$.} This indicates a considerable gain from a known
distribution function of output variables.

In order to visualize the dependence of the confidence level
$\beta$ on probability content $\gamma < \gamma_{cr}$ three curves
are shown in Fig. $14$ when $N=100$. The two upper curves
correspond to the known bivariate normal distribution of $y_{1}$
and $y_{2}$ with $\lambda = 2.5$, while the curve denoted by DF
refers to the distribution free case.

\section{Safety Inference}

The purpose of performing safety analysis is to assure that the
designed equipment can be operated safely. It is a
self-understanding premise that by altering input data randomly
within their prescribed distribution all the states will be either
safe or unsafe. If both safe and unsafe states would occur, the
entire range under consideration should be regarded as unsafe.

Our approach has severe consequences on every statement concerning
safety. The present section assesses those consequences. The first
consequence is that we can not speak of safety of a given state,
rather we can speak of the probability of a given state to be
safe. Assume for the sake of simplicity that the model is not
chaotic around the nominal state $\left(\vec{x}_0, \vec{y}_0
\right)$ where $\vec{y}_0 = \hat{\mbox C} \vec {x}_0$. The input
variable(s) may take values in a given range, that range is mapped
into a range of the output variables. From some other
considerations, which thus far have not been regarded as part of
safety analysis, we get information on the probability
distribution of the input variable. And, we select a range into
which a large portion, say more than 90{\%}, of the possible area
lies with a given high, say 95{\%}, probability. Consequently, we
conclude that:
\begin{enumerate}
\item It is insufficient to show that the nominal state is safe
because there may be probable inputs, which are unsafe. Therefore,
when the calculations are carried out exclusively in the nominal
state, safety analysis should demonstrate the estimated error to
be realistic.

\item Another possibility is that safety analysis should show that
images of all $x$ points in the vicinity of $x_{0}$ are safe. In
this case we get rid of the uncertainty caused by input
uncertainty.

\item Assumptions or knowledge of probability distributions of the
input do have an impact on safety issues, therefore they must not
be treated separately. Here two problems occur. Engineering input
data are usually not accompanied by probability distributions and
input variables, which are actually internal in the given
calculational model, may influence the output error decisively.
Such internal inputs are usually obtained from a fitting but that
procedure usually gives no information on the probability
distribution of fitted parameters (although theory and technique
are known).

\item Even if every $\hat{\mathcal C}\vec{x}$ is safe for a given
interval, there is a slight chance that some input(s) may be
associated with unsafe output(s). Those chances can be read out
from Table XI. for normally distributed output and from Tables IX.
and X. for a single arbitrarily distributed output variable, as
well as from Table XII. for two and three arbitrarily distributed
output variables.

\item Safety is described by random variables therefore we can
make only statistical assertions. Any claim concerning safety is
associated with ($\beta,\gamma)$ and the assumptions on the
probability distribution(s) of the input variables. An alarming
example is given in sub-subsection $3.2.3$ where we see that
$22.4$\% of the rejected calculations result in larger maxima than
in the basic sample.

\item Biased probability density functions seem to be extremely
dangerous. The simple problem of determining a quantile (see Fig.
$4$) may lead to large differences. Based on the presented
examples it seems desirable to treat certain class of
distributions of the output variables with special care.

\item Safety analysis should make it clear that every output
interval lies inside the safety envelope. The safety is not
unconditional but the values $\beta $ and $\gamma $ characterize
that "level" of safety. When any of inequalities $y_{k}(N) \geq
U_{T}^{(k)}, \; k=1, \ldots, n$ would be observed, then the system
operation could hardly be declared safe. This clearly indicates
that safety is not deterministic, as treated by many, but random.
\end{enumerate}

Since the general consideration results in loss of a large amount
of information, efforts should be made to chose a reasonable test
for the estimation of the probability distribution of the output
variable(s). To this end specific safety analysis models should be
analyzed individually. All these caveats necessitate a
reconsideration of safety issues.

\section{Conclusions}

The object of our investigation has been a complex system (e.g. a
computer code) that we treated as a black box: From a well-defined
input set the system (code) produces a well-defined output set.
Both sets have metrics; we can speak of distance between two input
sets or between two output sets. The computer code simulating the
complex system is a map $\hat{\mathcal C}: {\mathcal X} \to
{\mathcal Y}$, where ${\mathcal X}$ is the input set and
${\mathcal Y}$ is the output set. When analyzing given equipment,
we have a nominal input $x_{0}$ but the actual input might as well
be anywhere in ${\mathcal X}$, hence the input is a random vector.
The probability distribution of input components is usually
derived from diverse engineering considerations. In that setting,
we have to predict the {\em statistical behavior of the output
vector}; and, we have to specify a safety envelope into which the
actual output falls with high probability.~\footnote{ The present
work was initiated by a remark stating that a limited number of
runs suffices to determine the safety envelop.} The exact
statements are formulated as theorems in Sections $3$ and $4$,
while  the conclusions are summarized as follows.

\begin{enumerate}
\item The nominal state $\left(\vec{x}_0, \hat{\mathcal C} \vec
{x}_0 \right)$ is determined from the expectation value of the
input and from the associated output. The investigation of the
nominal state alone is insufficient to declare the system
operation to be safe.

\item When the distribution of output is not known, four methods,
namely the Bayesian, the percentile and the sign test as well as
the tolerance interval methods are proposed for testing the output
data. The statistical statements which can be obtained by these
methods do not differ significantly from one another. As expected,
since the distribution is  unknown, only a fraction of the
information present in the output can be utilized. As a result,
more runs are needed or lower probabilities are achieved.

\item When the output is of normal distribution, Theorem $3$
determines an interval $[L,U]$ around the estimated mean value of
the output into which a larger than a prescribed fraction $\gamma$
of the distribution falls with preassigned  probability $\beta$.
The limits $L$ and $U$ are determined by the sample estimate of
the standard deviation and by a positive factor $\lambda$. For the
mostly used $N, \beta, \gamma$ values the $\lambda$ factors are
given in Table XI. Our results are in accordance with Ref.
\cite{krzy90} and \cite{odeh80}.

\item When the output consists of more then one {\em statistically
not independent quantities}, the portion of information content
that can be utilized rapidly decreases with the number of
simultaneously tested output variables. That manifests again in a
larger number of runs or in lower probabilities. This is true for
both the sign test and tolerance interval methods. The results are
given in Tables IX. and XII. It is worth noting that our results
comply with the results given in Ref. \cite{krzy90} only when the
output variables are independent. To achieve identical
($\beta,\gamma$) level for statistically dependent outputs we need
a larger number of runs than given in Ref. \cite{krzy90}.
\end{enumerate}

All these observations may influence, in safety analysis, the
application of best estimate methods, and underline the opinion
that any realistic modeling and simulation of complex systems must
include the probabilistic features of the system and the
environment.

\renewcommand{\thesection}{\Roman{section}}
\setcounter{section}{0} \numberwithin{equation}{section}
\renewcommand{\theequation}{\mbox{I-\alph{equation}}}
\setcounter{equation}{0}

\section{Appendix. Proof of Theorem 1.}

Obvious, if $G(y)$ is continuous strictly increasing function of
$y$, then
\[ {\mathcal P}\{y(r) \leq Q_{\gamma} \leq y(s)\} =
{\mathcal P}\{y(r) \leq G^{-1}(\gamma) \leq y(s)\} = \]
\[ {\mathcal P}\{z(r) \leq \gamma \leq z(s),\}\]
which is nothing else than
\[ {\mathcal P}\{z(r) \leq \gamma \leq z(s),\} =
{\mathcal P}\{z(r) \leq \gamma, z(s) \geq \gamma,\}. \]
Introducing the notations ${\mathcal A} = \{z(s) \leq \gamma\}$
and $\overline{{\mathcal A}} = \{z(s) \geq \gamma\}$, we can write
that
\[ {\mathcal P}\{z(r) \leq \gamma\} = {\mathcal P}\{z(r) \leq
\gamma, {\mathcal A}+\overline{{\mathcal A}}\} = \]
\[ = {\mathcal P}\{z(r) \leq \gamma, z(s) \leq \gamma\} +
{\mathcal P}\{z(r) \leq \gamma, z(s) \geq> \gamma\}, \] and from
this we obtain
\[ {\mathcal P}\{y(r) \leq Q_{\gamma} \leq y(s)\} =
{\mathcal P}\{z(r) \leq \gamma, z(s) \geq \gamma\} = \]
\begin{equation} \label{a1}
 = {\mathcal P}\{z(r) \leq \gamma\} - {\mathcal P}\{z(r) \leq
\gamma, z(s) \leq \gamma\}.
\end{equation}

By using the well known expression
\[ {\mathcal P}\{u \leq z(r) \leq u+du, v \leq z(s)\leq v+dv\} =
g_{r,s}(u, v)\;du\;dv = \]
\begin{equation} \label{a2} =
\frac{u^{r-1}\;(v-u)^{s-r-1}\;(1-v)^{N-s}}{B(r, s-r)\; B(s,
N-s+1)}\;du\;dv,
\end{equation}
\[ 0 \leq u \leq v \leq 1,  \] we obtain that
\begin{equation} \label{a3}
\beta =  \int_{0}^{\gamma}\;\int_{0}^{1} g_{r,s}(u, v)\;du\;dv -
\int_{0}^{\gamma}\;\int_{0}^{\gamma} g_{r,s}(u, v)\;du\;dv.
\end{equation}
The first integral:
\[ Y_{1} = \int_{0}^{\gamma}\;\int_{0}^{1} g_{r,s}(u, v)\;du\;dv =
\int_{0}^{\gamma}g_{r}(u)\;du, \]
where
\[ g_{r}(u) = \frac{u^{r-1}\;(1-u)^{N-r}}{B(r, N-r+1)}, \]
hence
\[ Y_{1} = I(\gamma, r, N-r+1). \]
Taking into account that $v \geq u$ the second integral is nothing
else than
\[ Y_{2} = \int_{0}^{\gamma}\;\int_{0}^{\gamma} g_{r,s}(u, v)\;du\;dv = \]
\[ = C_{r,s}\;\int_{0}^{\gamma} dv \int_{0}^{v} u^{r-1}\;
(v-u)^{s-r-1}\;(1-v)^{N-s}\;du, \] where
\begin{equation} \label{a4}
C_{r,s} = \frac{1}{B(r, s-r)\;B(s, N-s+1)}.
\end{equation}
By performing the transformations
\[ u = t_{1}t_{2} \;\;\;\;\;\; \mbox{and} \;\;\;\;\;\; v = t_{2}\]
the integral $Y_{2}$ can be easily calculated. Since
\[ J = \left|\begin{array}{ll}
\frac{\partial{u}}{\partial t_{1}} & \frac{\partial{u}}{\partial
t_{2}} \\ \mbox{ } & \mbox{ } \\ \frac{\partial{v}}{\partial
t_{1}} & \frac{\partial{v}}{\partial t_{2}}
\end{array} \right| = t_{2}, \]
we find that
\[ Y_{2} = C_{r,s}\;\int_{0}^{\gamma} t_{2}\; dt_{2} \int_{0}^{1}(t_{1}
t_{2})^{r-1}\;(t_{2}-t_{1} t_{2})^{s-r-1}\;(1-t_{2})^{N-s}\;dt_{1}
= \] \[ =
C_{r,s}\;\int_{0}^{1}t_{1}^{r-1}\;(1-t_{1})^{s-r-1}\;dt_{1}\;
\int_{0}^{\gamma} t_{2}^{s-1}\;(1-t_{2})^{N-s}\;dt_{2}. \]
Replacing $C_{r,s}$ by (\ref{a3}) one obtains that
\[ Y_{2} = I(\gamma, s, N-s+1). \]  By using the well known identity
$I(c, a, b) = 1 - I(1-c, b, a)$, finally we have
\[ \beta = {\mathcal P}\{y(r) \leq Q_{\gamma} \leq y(s)\} =
Y_{1} - Y_{2} = \]
\[ = I(1-\gamma, N-s+1, s) - I(1-\gamma, N-r+1, r),
\] and so the Theorem 1 is proven.

\renewcommand{\theequation}{\mbox{II-\alph{equation}}}
\setcounter{equation}{0}

\section{Appendix. Proof of Theorem 2.}

The derivation of Eq. $(22)$ is based on the following
observation. Let
\[ G(y) = \int_{-\infty}^{y} g(t)\;dt \]
be the unknown but continuous cumulative distribution function of
the output variable $y$. Let $y_{1}, \ldots, y_{N}$ be its
independently observed values. Arrange the sample elements $y_{k},
\;\; k=1, \ldots, N$ in increasing order, and denote by $y(k)$ the
$k$th element of the ordered sample. Introduce the random
variables $z(k) = G[y(k)], \;\; k = 1, \ldots, N$ which are are
not independent \cite{lpal95}. According to (\ref{a2}) the
bivariate density function of $z(s)$ and $z(r), \;\; r < s$ is
given by
\[ g_{(r,s)}^{(N)}(u, v) =
\frac{N!}{(r-1)!\;(s-r+1)!\;(N-s)!}\;u^{r-1}\;(v-u)^{r-s+1}\;v^{N-s},
\]
\[ 0 \leq u \leq v \leq 1. \]
In order to determine the probability
\[ {\mathcal P}\left\{\int_{y(r)}^{y(s)} dG(y) >
\gamma\right\} = \] \[ = {\mathcal P}\left\{G[y(s)] - G[y(r)] >
\gamma\right\} = {\mathcal P}\left\{z(s) - z(r)] > \gamma\right\},
\] we need the probability
\begin{equation} \label{b1}
{\mathcal P}\left\{t \leq z(s) - z(r) \leq t+dt\right\} =
w_{(r,s)}^{(N)}(t)\;dt = \int_{0}^{1-t}g_{(r,s)}^{(N)}(u,
u+t)\;du\;dt.
\end{equation}
Substituting $g_{r,s}^{(N)}(u,v)$ into (\ref{b1}), the integration
in (\ref{b1}) can be carried out:
\begin{equation} \label{b2}
w_{(r,s)}^{(N)}(t) = \frac{1}{B(r, s-r)\;B(s, N-s+1)}\;t^{s-r+1}
\int_{0}^{1-t} u^{r-1}\;(1-t-u)^{N-s}\;du,
\end{equation}
where $B(j,k)$ is the Euler beta function. Taking this expression
into account, we get
\[{\mathcal P}\left\{z(s) - z(r) > \gamma\right\} =
\int_{\gamma}^{1}w_{(r,s)}^{(N)}(t)\;dt, \] after integration we
obtain
\[ {\mathcal P}\left\{z(s) - z(r) > \gamma\right\} = 1 - \int_{0}^{\gamma}
\frac{t^{s-r-1}\;(1-t)^{N-s-r}}{B(s-r, N-s+r+1)}\;dt. \] In other
words,
\begin{equation} \label{b3}
\beta = 1 - I(\gamma, s-r, N-s+r+1)
\end{equation}
as stated. Q.E.D.

\renewcommand{\theequation}{\mbox{III-\alph{equation}}}
\setcounter{equation}{0}

\section{Appendix. Proof of Theorem 3.}

The proof of the Theorem 3 is based on a few well known relations
of mathematical statistics. By the definition of conditional
probability,
\begin{equation}  \label{c1}
{\mathcal P}\{\rho(\tilde{z}, \tilde{s}) > \gamma\} =
\int_{-\infty}^{+\infty}{\mathcal P}\{\rho(\tilde{z}, \tilde{s}) >
\gamma|\tilde{z}=\mu\}\;d{\mathcal P}\{\tilde{z} \leq \mu\},
\end{equation}
where
\[ d{\mathcal P}\{\tilde{z} \leq \mu\} = d{\mathcal
P}\left\{\frac{1}{N}\sum_{n=1}^{N} \frac{y_{n}-m}{\sigma} \;\leq
\; \mu\right\} = \sqrt{\frac{N}{2\pi}}\;e^{-N\mu^{2}/2}\;d\mu. \]
Since
\[ {\mathcal P}\{\rho(\tilde{z}, \tilde{s}) >
\gamma|\tilde{z}=\mu\}= {\mathcal P}\left\{\rho(\mu, \tilde{s}) >
\gamma\right\}, \] we have
\begin{equation} \label{c2}
{\mathcal P}\{\rho(\tilde{z}, \tilde{s}) > \gamma\} =
\sqrt{\frac{N}{2\pi}}\;\int_{-\infty}^{+\infty}{\mathcal
P}\left\{\rho(\mu, \tilde{s}) >
\gamma\right\}\;e^{-N\mu^{2}/2}\;d\mu.
\end{equation}
where
\[ \rho(\mu, \tilde{s}) = \frac{1}{\sqrt{2\pi}}\;\int_{\mu -
\lambda\tilde{s}}^{\mu + \lambda\tilde{s}} e^{-z^{2}/2}\;dz \] is
random variable. Let us define the function
\[ r(\mu, \lambda s) = \frac{1}{\sqrt{2\pi}}\;\int_{\mu -
\lambda s}^{\mu + \lambda s} e^{-z^{2}/2}\;dz \] for real $s$. If
$\mu$ and $\lambda$ are fixed, then $r(\mu, \lambda s)$ is
strictly monotonously increasing function of $s$, therefore the
equation
\[ \frac{1}{\sqrt{2\pi}}\;\int_{\mu -
\lambda s}^{\mu + \lambda s} e^{-z^{2}/2}\;dz = \gamma \] has only
one root in $s$. It is clear that $\lambda s$ is independent of $
\lambda$, hence we may write $\lambda s = q(\mu, \gamma)$ and $q$
is obtained from
\[ \frac{1}{\sqrt{2\pi}}\;\int_{\mu - q}^{\mu + q} e^{-z^{2}/2}\;dz =
\gamma. \] It follows from the property of $r(\mu, \lambda s)$
that the probability of $\rho(\mu, \tilde{s}) > \gamma$ equals to
the probability of $\tilde{s} > s_{r}$, i.e.
\begin{equation} \label{c3}
{\mathcal P}\left\{\rho(\mu, \tilde{s}) > \gamma\right\} =
{\mathcal P}\left\{\tilde{s} > s_{r}\right\} = {\mathcal
P}\left\{\tilde{s}
> \frac{q(\mu, \gamma)}{\lambda}\right\}.
\end{equation}
By using this relations we can write that
\[ {\mathcal P}\left\{\tilde{s} > \frac{q(\mu, \gamma)}{\lambda}\right\} =
{\mathcal P}\left\{\tilde{s}^{2}
> \frac{[q(\mu, \gamma)]^{2}}{\lambda^{2}}\right\} = {\mathcal
P}\left\{\frac{\tilde{\sigma}^{2}}{\sigma^{2}}
> \frac{[q(\mu, \gamma)]^{2}}{\lambda^{2}}\right\}, \]
and taking into account that the random variable
\[ (N-1)\;\frac{\tilde{\sigma}^{2}}{\sigma^{2}} = \sum_{n=0}^{N}
\left(\frac{y_{n}-\tilde{y}}{\sigma}\right)^{2} \] is of
$\chi^{2}$ distribution with $(N-1)$ degree of freedom
\cite{lpal95}, we get
\begin{equation} \label{c4}
{\mathcal P}\left\{\rho(\mu, \tilde{s}) > \gamma\right\} = 1 -
K_{N-1}\left\{(N-1)\;\frac{[q(\mu, \gamma)]^{2}}
{\lambda^{2}}\right\},
\end{equation}
where
\[ K_{N}(x) = \frac{1}{2\Gamma(N/2)}\;\int_{0}^{x}
\left(\frac{u}{2}\right)^{(N-2)/2}\;e^{-u/2}\;du. \] Substituting
(\ref{c4}) into (\ref{c2}) we get the theorem proven. Q.E.D.

\renewcommand{\theequation}{\mbox{IV-\alph{equation}}}
\setcounter{equation}{0}

\section{Appendix. Proof of Theorem 4.}

Before setting out for the proof of Theorem 4, we set forth the
following notation. Let
\[ H(\lambda, \gamma|\mu) = {\mathcal P}\left\{\rho(\mu,
\tilde{s}) > \gamma\right\}. \] We need
\begin{lemma}
It can be shown that
\begin{equation} \label{d1}
H(\lambda, \gamma|1/\sqrt{N}) - W(\lambda, \gamma, N) = O(N^{-2}).
\end{equation}
where
\begin{equation} \label{d2}
W(\lambda, \gamma, N) =
\sqrt{\frac{N}{2\pi}}\;\int_{-\infty}^{+\infty} H(\lambda,
\gamma|\mu) \;e^{-\mu^{2}/2}\;d\mu.
\end{equation}
\end{lemma}

\vspace{0.3cm}

{\em Proof of Lemma 1}. The expression
\begin{equation} \label{d3}
H(\lambda, \gamma|\mu) = {\mathcal P}\left\{\frac{1}{\sqrt{2\pi}}
\;\int_{\mu-\lambda \tilde{s}}^{\mu+\lambda \tilde{s}}
e^{-z^{2}/2}\;dz > \gamma\right\}
\end{equation}
is an even function of $\mu$ and can be developed into Taylor
series around $\mu=0$ as
\begin{equation} \label{d4}
H(\lambda, \gamma|\mu) = \sum_{n=0}^{\infty}
\left[\frac{\partial^{2n}H(\lambda, \gamma|\mu)}{\partial
\mu^{2n}}\right]_{\mu=0}\;\frac{\mu^{2n}}{(2n)!}.
\end{equation}
Substituting (\ref{d4}) into (\ref{d3}) we obtain
\[ W(\lambda, \gamma, \mu) = \sum_{n=0}^{\infty}
\left[\frac{\partial^{2n}H(\lambda, \gamma|\mu)}{\partial
\mu^{2n}}\right]_{\mu=0}\;\frac{(2n-1)!!}{(2n)!}\frac{1}{N^{n}} =
\]
\begin{equation} \label{d5}
=  H(\lambda, \gamma|0) + \left[\frac{\partial^{2}H(\lambda,
\gamma|\mu)}{\partial \mu^{2}}\right]_{\mu=0}\;\frac{1}{2N} +
O(N^{-1/2}),
\end{equation}
and replacing $\mu$ by $1/\sqrt{N}$ in (\ref{d4}), we have
\begin{equation} \label{d6}
H(\lambda, \gamma|1/\sqrt{N}) = H(\lambda, \gamma|0) +
\left[\frac{\partial^{2}H(\lambda, \gamma|\mu)}{\partial
\mu^{2}}\right]_{\mu=0}\;\frac{1}{2N} + O(N^{-1/2}).
\end{equation}
From Eqs. (\ref{d6}) and (\ref{d5}) follows that (\ref{d1}) is
true. Consequently , we have the following approximate equation
\[ H(\lambda, \gamma|1/\sqrt{N}) = 1 -  K_{N-1}\left[(N-1)\;\frac{[q(N^{-1/2},
\gamma)]^{2}}{\lambda^{2}}\right] \approx  \beta, \] where the
argument of $K_{N-1}[\cdots]$ is nothing else than the $(1-\beta)$
percentile of $\chi^{2}$ distribution with $(N-1)$ degree of
freedom. Introducing the notation
\[ Q_{N-1}(1-\beta) = (N-1)\;\frac{[q(N^{-1/2},
\gamma)]^{2}}{\lambda^{2}},   \] we find that
\begin{equation} \label{d7}
\lambda \approx \lambda_{a}(\gamma, \beta) =
\sqrt{\frac{N-1}{Q_{N-1}(1-\beta)}}\;q(1/\sqrt{N},\gamma).
\end{equation}
This completes the proof of Theorem $2$. Q.E.D.

\renewcommand{\theequation}{\mbox{V-\alph{equation}}}
\setcounter{equation}{0}

\section{Appendix. Proof of Theorem 5.}

The proof of Theorem 5 is given in two steps. In the first step we
show that the Theorem holds for $n=2$, and then we generalize the
claim for $n>2$.

{\em Step 1.} We assume that the unknown joint distribution
function of two output variables $y_{1}$ and $y_{2}$ is given by
\[ G(y_{1}, y_{2}) = \int_{-\infty}^{y_{1}}\;\int_{-\infty}^{y_{2}}
g(t_{1}, t_{2})\;dt_{1}\;dt_{2}, \] and denote by
\[ g_{1}(y_{1}) = \int_{-\infty}^{+\infty} g(y_{1},
t_{2})\;dt_{2} \] the density function of the output variable
$y_{1}$. Let us consider the following random variable
\begin{equation} \label{e1}
{\mathcal A}_{2} = {\mathcal A}_{2}(L_{1}, U_{1}, L_{2}, U_{2}) =
\int_{L_{1}}^{U_{1}}\;\int_{L_{2}}^{U_{2}} g(y_{1}, y_{2})\;dy_{1}
dy_{2},
\end{equation}
where the boundaries of the integration are random variables. The
limits were discussed in Section 4.2.2. ${\mathcal A}_{2}$ can be
expressed almost surely as
\begin{equation} \label{e2}
{\mathcal A}_{2}(L_{1}, U_{1}, L_{2}, U_{2}) = {\mathcal
C}_{2}(L_{2}, U_{2}|L_{1}, U_{1})\;{\mathcal A}_{1}(L_{1}, U_{1}).
\end{equation}
Here
\begin{equation} \label{e3}
{\mathcal A}_{1}(L_{1}, U_{1}) =
\int_{L_{1}}^{U_{1}}g_{1}(y_{1})\;dy_{1}
\end{equation}
and
\begin{equation} \label{e4}
{\mathcal C}_{2}(L_{2}, U_{2}|L_{1}, U_{1}) =
\int_{L_{2}}^{U_{2}}\phi_{2}(y_{2}|L_{1},U_{1})\;dy_{2}.
\end{equation}
where
\begin{equation} \label{e5}
\phi_{2}(y_{2}|L_{1},U_{1}) = \frac{\int_{L_{1}}^{U_{1}} g(y_{1},
y_{2})\;dy_{1}}{\int_{-\infty}^{+\infty}dy_{2}\;\int_{L_{1}}^{U_{1}}
g(y_{1}, y_{2})\;dy_{1}} = \frac{\int_{L_{1}}^{U_{1}} g(y_{1},
y_{2})\;dy_{1}}{{\mathcal A}_{1}(L_{1}, U_{1})}
\end{equation}
is the random density of variable $y_{2}$ under the condition that
$y_{1}$ lies in $[L_{1},U_{1}]$. Since ${\mathcal A}_1(L_1 ,U_1) =
G_1[y_1(s_1)] - G_1[y_1(r_1)]$, using relation (IV-b), we find
that
\[ {\mathcal P}\left\{t_{1} \leq {\mathcal A}_{1}(L_{1}, U_{1}) \leq
t_{1} + dt_{1}\right\} =
\frac{t_{1}^{s_{1}-r_{1}-1}\;(1-t_{1})^{N-s_{1}+r_{1}}}{B(s_{1}-r_{1},
N-s_{1}+r_{1}+1)}\;dt_{1} = \]
\begin{equation} \label{e6}
= k_{(r_{1},s_{1})}^{(N)}(t_{1})\;dt_{1}.
\end{equation}
To obtain the density function of ${\mathcal C}_{2}(L_{2},
U_{2}|L_{1}, U_{1})$, we define the random probability measure
\[ G(t|L_{1},U_{1}) =
\int_{-\infty}^{t}\phi_{2}(y_{2}|L_{1},U_{1})\;dy_{2}, \] with
which we can express ${\mathcal C}_{2}$ as
\[ {\mathcal C}_{2}(L_{2}, U_{2}|L_{1}, U_{1}) =
G[y_{2}(s_{2})|L_{1},U_{1}] - G[y_{2}(r_{2})|L_{1},U_{1}],  \]
where $r_{1} \leq r_{2} < \cdots < s_{2} \leq s_{1}$.  Finally we
get
\[{\mathcal P}\left\{ t_{2} \leq {\mathcal
C}_{2}(L_{2}, U_{2}|L_{1}, U_{1}) \leq t_{2} + dt_{2}\right\} = \]
\begin{equation} \label{e7}
= \frac{t_{2}^{s_{2}-r_{2}-1}\;
(1-t_{2})^{s_{1}-r_{1}-1-s_{2}+r_{2}}} {B(s_{2}-r_{2},
s_{1}-r_{1}-s_{2}+r_{2})}\;dt_{2} =
\ell_{(r_{2},s_{2})}^{(s_{1}-r_{1}-1)}(t_{2})\;dt_{2}.
\end{equation}

\vspace{0.2cm}

Note that expression (\ref{e7}) contains neither $L_{1}$ nor
$U_{1}$, therefore, distribution of random variable ${\mathcal
C}_{2}$ is independent of $L_{1}$ and $U_{1}$. Consequently, the
joint density distribution of ${\mathcal A}_{1}$ and ${\mathcal
C}_{2}$ is the product of (\ref{e6}) and (\ref{e7}). We still need
the density function of the random variable ${\mathcal A}_{2}$.
Exploiting the independence of ${\mathcal C}_{2}$ and ${\mathcal
A}_{1}$, we get

\newpage

\[ {\mathcal P}\left\{ t \leq {\mathcal A}_{2}(L_{1}, U_{1}, L_{2},
U_{2}) \leq t + dt\right\} = \]
\begin{equation} \label{e8}
= \int_{t}^{1}\frac{1}{x}\;k_{(r_{1},s_{1})}^{(N)}(x)\;
\ell_{(r_{2},s_{2})}^{(s_{1}-r_{1}-1)}(t/x)\;dx\;dt = w_{{\mathcal
A}_{2}}(t)\;dt.
\end{equation}
Substituting here (\ref{e6}) and (\ref{e7}) and performing the
indicated calculations we obtain:
\begin{equation} \label{e9}
w_{{\mathcal A}_{2}}(t) =
\frac{t^{s_{2}-r_{2}-1}\;(1-t)^{N-s_{2}+r_{2}}}{B(s_{2}-r_{2},
N-s_{2}+r_{2}+1)}.
\end{equation}
From this, immediately follows
\[ {\mathcal P}\left\{{\mathcal A}_{2}(L_{1}, U_{1}, L_{2},
U_{2}) > \gamma\right\} = \] \[ = 1 - \frac{B(\gamma, s_{2}-r_{2},
N-s_{2}+r_{2}+1)}{B(s_{2}-r_{2}, N-s_{2}+r_{2}+1)} = 1 -
I(\gamma,s_{2}-r_{2}, N-s_{2}+r_{2}+1). \] This completes Step 1.

{\em Step 2.} Now we generalize the above result for $n > 2$. Let
us assume that the unknown joint probability distribution of the
output variables $y_{1}, \ldots, y_{n}$ is given by
\[ G(y_{1}, \ldots, y_{n}) = \int_{-\infty}^{y_{n}} \cdots
\int_{-infty}^{Y_{1}} g(v_{1}, \ldots, v_{n})\; dv_{1} \cdots
dv_{n}. \] Our task is to derive the probability distribution  of
the random variable
\[{\mathcal A}_{p}\left(L_{1}, U_{1}, \ldots,
L_{n}, U_{n}\right) = \int_{L_{n}}^{U_{n}} \cdots
\int_{L_{1}}^{U_{1}} g(y_{1}, \ldots, y_{n})\; dy_{1} \cdots
dy_{n}, \]  which is an $n$-fold integral over the $n$ output
variables. We introduce an intermediate term, in which an $i$-fold
definite integral over the first $i$ variables is involved, and
the rest of the variables are integrated over the $[-\infty,
+\infty]$ range:
\[ {\mathcal A}_{i}\left(L_{1}, U_{1}, \ldots,
L_{i}, U_{i}\right) = \] \[ = \int_{-\infty}^{+\infty} dy_{n}
\cdots \int_{-\infty}^{+\infty} dy_{i+1}\; \int_{L_{i}}^{U_{i}}
dy_{i} \cdots \int_{L_{1}}^{U_{1}} dy_{1}\; g(y_{1}, \ldots,
y_{n}), \] and
\[ \phi_{i}\left(y_{i}|L_{1}, U_{1}, \ldots, L_{i-1},
U_{i-1}\right) = \frac{1}{{\mathcal
A}_{i-1}}\;\int_{L_{i-1}}^{U_{i-1}}dy_{i-1} \cdots
\int_{L_{1}}^{U_{1}} dy_{1}\; g(y_{1}, \ldots, y_{n}), \] which is
the random density of the variable $y_{i}$ under the condition
that $L_{j} \leq y_{j} \leq U_{j}, \;\; j=1, \ldots, i-1$. As we
did in (\ref{e4}), we introduce a random probability measure
associated with the condition that the first $(i-1)$ output
variables lie in the interval assigned to them by $[L_{j},
U_{j}],\;\; j=1,\ldots,i-1$:
\[ {\mathcal C}_{i} = {\mathcal C}_{i}\left(L_{i}, U_{i}|L_{1},
U_{1}, \ldots, L_{i-1}, U_{i-1}\right) =
\int_{L_{i}}^{U_{i}}\phi_{i}\left(y_{i}|L_{1}, U_{1}, \ldots,
L_{i-1}, U_{i-1}\right)\;dy_{i}. \] The above defined ${\mathcal
A}_{i}$'s obey the recursion
\begin{equation} \label{e10}
{\mathcal A}_{i+1} = {\mathcal C}_{i+1}\;{\mathcal A}_{i}.
\end{equation}

\begin{lemma}
The probability of finding ${\mathcal A}_{i}$ in the interval
$[t_{i}, t_{i}+dt_{i}]$ is given by
\begin{equation} \label{e11}
{\mathcal P}\left\{t_{i} \leq {\mathcal A}_{i} \leq
t_{i}+dt_{i}\right\} = \frac{t_{i}^{s_{i}-r_{i}-1}\;(1 -
t_{i})^{N-s_{i}+r_{i}}}{B(s_{i}-r_{i}, N-s_{i}+r_{i}+1)}\;dt_{i}.
\end{equation}
\end{lemma}

\vspace{0.3cm}

{\em Proof of Lemma 2.} Eq. (\ref{e11}) is certainly true for
$i=1, 2$ because
\[ {\mathcal A}_{1} = {\mathcal A}_{0}\;{\mathcal C}_{1} =
{\mathcal C}_{1} = \int_{L_{1}}^{U_{1}} g(y)\;dy \] and
\[ {\mathcal A}_{2} = {\mathcal C}_{2}\;{\mathcal A}_{1} =
{\mathcal C}_{2}(L_{2}, U_{2}|L_{1}, U_{1})\;{\mathcal
A}_{1}(L_{1}, U_{1}) = \int_{L_{2}}^{U_{2}}\;\int_{L_{1}}^{U_{1}}
g(y_{1}, y_{2})\;dy_{1}\;dy_{2}. \] Now we assume that (\ref{e11})
is true for $i=j$ and show that it is true also for $i=j+1$. Note
that ${\mathcal A}_{j}$ and ${\mathcal C}_{j+1}$ are statistically
independent because
\[ {\mathcal P}\left\{t_{j+1} \leq {\mathcal C}_{j+1} \leq
t_{j+1}+dt_{j+1}\right\} = \]
\[ = \frac{t_{j+1}^{s_{j+1}-r_{j+1}-1}\;(1 -
t_{j+1})^{s_{j}-r_{j}-1-s_{j+1}+r_{j+1}}}{B(s_{j+1}-r_{j+1},
s_{j}-r_{j}-s_{j+1}+r_{j+1})}\;dt_{j+1} \] does not involve the
quantities $L_{j}, U_{j}, \ldots, L_{1}, U_{1}$, which occur in
${\mathcal A}_{j}$. The joint density function of ${\mathcal
A}_{j}$ and ${\mathcal C}_{j+1}$ takes the form of the joint
density function of ${\mathcal A}_{1}$ and ${\mathcal C}_{2}$ in
the case of $n=2$. Hence the density function of
\[ {\mathcal A}_{j}\;{\mathcal C}_{j+1} = {\mathcal A}_{j+1} \]
is obtainable from Eq. (\ref{e9}) by substituting $r_{j+1}$ for
$r_{2}$ and $s_{j+1}$ for $s_{2}$, i.e.
\[ {\mathcal P}\left\{t_{j+1} \leq {\mathcal A}_{j+1} \leq
t_{j+1}+dt_{j+1}\right\} = \]
\[ = \frac{t_{j+1}^{s_{j+1}-r_{j+1}-1}\;(1 -
t_{j+1})^{N-s_{j+1}+r_{j+1}}}{B(s_{j+1}-r_{j+1},
N-s_{j+1}+r_{j+1}+1)}\;dt_{j+1}. \] Hence, Eq. (\ref{e11}) is
proven for $i = 1, 2, \ldots, n$. This completes Step 2.
Furthermore, the density function of ${\mathcal A}_{n}$ is given
by
\[ {\mathcal P}\left\{t \leq {\mathcal A}_{n} \leq
t+dt\right\} = w_{{\mathcal A}_{n}}(t)\;dt =
\frac{t^{s_{n}-r_{n}-1}\;(1 - t)^{N-s_{n}+r_{n}}}{B(s_{n}-r_{n},
N-s_{n}+r_{n}+1)}\;dt. \] It is interesting to note that the
density function of ${\mathcal A}_{n}$ does not depend on the
integers $r_{1}, s_{1}, \ldots, r_{n-1}, s_{n-1}$. This completes
the proof of Theorem $4$. Q.E.D.

\end{document}